\documentclass[acmsmall,screen,nonacm]{acmart}

\newif\ifarxiv
\arxivtrue

\AtBeginDocument{%
  }

\newcommand{\AFL}{{\smaller AFL}\xspace}
\newcommand{\fandango}{{\smaller FANDANGO}\xspace}
\newcommand{\isla}{{\smaller ISL}a\xspace}
\newcommand{\libfuzzer}{{libFuzzer}\xspace}

\newcommand{\tool}{{\smaller FANDANGO-RS}\xspace}

\newcommand{\Zthree}{{Z3}\xspace}

\newcommand{\sizeConstraintMinStatements}{5\xspace}
\newcommand{\sizeConstraintStructLimit}{1\xspace}

\newcommand{\sizeConstraintFunLimit}{1\xspace}

\newcommand{\multiValidAndSizedKPathCoverageAvgRs}{68.87\%\xspace}
\newcommand{\multiValidAndSizedTotalAvgRs}{939\xspace}

\newcommand{\multiValidAndSizedCompileAvgRs}{877\xspace}
\newcommand{\multiValidAndSizedGenTimeAvgRs}{37.28s\xspace}
\newcommand{\multiValidAndSizedCompileTimeAvgRs}{25.11s\xspace}
\newcommand{\multiValidAndSizedTotalTimeAvgRs}{62.40s\xspace}
\newcommand{\multiValidKPathCoverageAvgRs}{69.43\%\xspace}
\newcommand{\multiValidTotalAvgRs}{2143\xspace}

\newcommand{\multiValidCompileAvgRs}{2142\xspace}
\newcommand{\multiValidGenTimeAvgRs}{0.54s\xspace}
\newcommand{\multiValidCompileTimeAvgRs}{61.21s\xspace}
\newcommand{\multiValidTotalTimeAvgRs}{61.75s\xspace}
\newcommand{\singleValidAndSizedKPathCoverageAvgRs}{0.00\%\xspace}
\newcommand{\singleValidAndSizedTotalAvgRs}{584\xspace}

\newcommand{\singleValidAndSizedCompileAvgRs}{0\xspace}
\newcommand{\singleValidAndSizedGenTimeAvgRs}{63.35s\xspace}
\newcommand{\singleValidAndSizedCompileTimeAvgRs}{0.00s\xspace}
\newcommand{\singleValidAndSizedTotalTimeAvgRs}{63.35s\xspace}
\newcommand{\singleValidKPathCoverageAvgRs}{71.87\%\xspace}
\newcommand{\singleValidTotalAvgRs}{2145\xspace}

\newcommand{\singleValidCompileAvgRs}{2145\xspace}
\newcommand{\singleValidGenTimeAvgRs}{0.29s\xspace}
\newcommand{\singleValidCompileTimeAvgRs}{60.80s\xspace}
\newcommand{\singleValidTotalTimeAvgRs}{61.09s\xspace}
\newcommand{\unconstrainedKPathCoverageAvgRs}{72.62\%\xspace}
\newcommand{\unconstrainedTotalAvgRs}{5600\xspace}

\newcommand{\unconstrainedCompileAvgRs}{656\xspace}
\newcommand{\unconstrainedGenTimeAvgRs}{0.02s\xspace}
\newcommand{\unconstrainedCompileTimeAvgRs}{60.69s\xspace}
\newcommand{\unconstrainedTotalTimeAvgRs}{60.71s\xspace}
\newcommand{\unconstrainedKPathCoverageAvgPy}{66.30\%\xspace}
\newcommand{\unconstrainedTotalAvgPy}{3100\xspace}

\newcommand{\unconstrainedCompileAvgPy}{1493\xspace}
\newcommand{\unconstrainedGenTimeAvgPy}{3.94s\xspace}
\newcommand{\unconstrainedCompileTimeAvgPy}{57.56s\xspace}
\newcommand{\unconstrainedTotalTimeAvgPy}{61.50s\xspace}
\newcommand{\validConstrainedKPathCoverageAvgPy}{51.73\%\xspace}
\newcommand{\validConstrainedTotalAvgPy}{540\xspace}

\newcommand{\validConstrainedCompileAvgPy}{540\xspace}
\newcommand{\validConstrainedGenTimeAvgPy}{48.93s\xspace}
\newcommand{\validConstrainedCompileTimeAvgPy}{15.84s\xspace}
\newcommand{\validConstrainedTotalTimeAvgPy}{64.77s\xspace}

\newcommand{\conclusion}[1]{%
  \begin{center}
    \fbox{\begin{minipage}{0.95\columnwidth}
      \centering
      \emph{#1}
    \end{minipage}}
  \end{center}
}

\def\itemrange#1{%
\addtocounter{enumi}{1}%
\edef\labelenumi{(\theenumi--\noexpand\theenumi)}%
\addtocounter{enumi}{-1}%
\addtocounter{enumi}{#1}%
\item
\def\labelenumi{(\theenumi)}}

\def\nonterm#1{\textnormal{$\langle$\emph{#1}$\rangle$}}

\usepackage[nounderscore]{syntax}\newenvironment{densegrammar}{%
\begin{small}
\grammarparsep 3pt
\parsep 0pt
\itemsep 0pt
\begin{grammar}
}%
{%
\end{grammar}
\end{small}
}

\usepackage{tikz}
\usetikzlibrary{automata,positioning,shapes.callouts}

\usepackage[frozencache=true,cachedir=.]{minted}

\usepackage{xpatch,letltxmacro}
\LetLtxMacro{\cminted}{\minted}

\xpretocmd{\cminted}{\RecustomVerbatimEnvironment{Verbatim}{BVerbatim}{}}{}{}

\usepackage{tabularx}
\usepackage{booktabs}
\usepackage[para]{threeparttable}
\usepackage{multirow}

\usepackage[capitalise, nameinlink]{cleveref}
\usepackage{multicol}
\usepackage{subcaption}
\usepackage{nicefrac,xfrac}
\usepackage{xspace}
\usepackage[final]{zebra-goodies}\usepackage[compatibility=false]{caption}
\usepackage[inline]{enumitem}
\usepackage{pifont}

\newcommand{\cmark}{\ding{51}}
\newcommand{\xmark}{\ding{55}}

\newcommand{\fandangogenfactor}{{3--4 orders of magnitude}}

\setcopyright{rightsretained}

\begin{document}

\title{Efficient Multi-Objective Evolutionary Test Generation}
\subtitle{Making Evolutionary Language-Based Testing Feasible with Extensible Generated Rust}

\title{High-Performance Generation of Constrained Inputs}
\subtitle{Optimizing Evolutionary Language-Based Testing with Grammar-to-Type Transpilation}

\author{Addison Crump}
\email{addison.crump@cispa.de}
\orcid{0009-0003-3271-3558}
\affiliation{%
  \institution{CISPA Helmholtz Center for Information Security}
  \country{Germany}
}

\author{Alexi Turcotte}
\email{alexi.turcotte@cispa.de}
\orcid{0000-0002-0381-0477}
\affiliation{%
  \institution{CISPA Helmholtz Center for Information Security}
  \country{Germany}
}

\author{Jos\'e Antonio Zamudio Amaya}
\email{jose.zamudio@cispa.de}
\orcid{0000-0002-5025-7424}
\affiliation{%
  \institution{CISPA Helmholtz Center for Information Security}
  \country{Germany}
}

\author{Andreas Zeller}
\email{andreas.zeller@cispa.de}
\orcid{0000-0003-4719-8803}
\affiliation{%
  \institution{CISPA Helmholtz Center for Information Security}
  \country{Germany}
}

\ccsdesc[500]{Software and its engineering~Software testing and debugging}
\ccsdesc[500]{Software and its engineering~Empirical software validation}
\ccsdesc[300]{Software and its engineering~Translator writing systems and compiler generators}
\ccsdesc[300]{Software and its engineering~Source code generation}
\ccsdesc[300]{Software and its engineering~Formal language definitions}
\ccsdesc[300]{Theory of computation~Grammars and context-free languages}
\ccsdesc[300]{Software and its engineering~Search-based software engineering}
\ccsdesc[300]{Theory of computation~Generating random combinatorial structures}
\ccsdesc[100]{Software and its engineering~Syntax}
\ccsdesc[100]{Software and its engineering~Semantics}
\ccsdesc[100]{Software and its engineering~Constraints}
\ccsdesc[100]{Theory of computation~Evolutionary algorithms}

\keywords{Rust, Transpiler, Type Safety, Source-to-Source, Language-based Testing}

\begin{abstract}
Language-based testing combines context-free grammar definitions with semantic constraints over grammar elements to generate test inputs.
By pairing context-free grammars with constraints, users have the expressiveness of unrestricted grammars while retaining a simple structure.
However, producing inputs in the presence of such constraints can be challenging.
In past approaches, SMT solvers have been found to be very slow at finding string solutions; evolutionary algorithms are faster and more general, but current implementations still struggle with complex constraints that would be required for domains such as compiler testing.

In this paper, we present a novel approach for evolutionary language-based testing that improves performance by \fandangogenfactor{} over the current state of the art, reducing hours of generation and constraint solving time to seconds.
We accomplish this by (1)~carefully transforming grammar definitions into Rust types and trait implementations, ensuring that the compiler may near-maximally optimize arbitrary operations on arbitrary grammars; and
(2)~using better evolutionary algorithms that improve the ability of language-based testing to solve complex constraint systems.
The synthesis of our performance and algorithmic improvements allow our prototype, \tool, to solve constraints that previous strategies simply cannot handle.
We demonstrate this by a case study for a C subset, in which \tool is able to generate \multiValidAndSizedCompileAvgRs diverse, complex, and valid test inputs for a C compiler per minute.

\end{abstract}

\maketitle

\section{Introduction} \label{sec:intro}

To systematically produce valid test inputs for a program, one must specify its \emph{language}.
A common specification formalism for such input languages is a \emph{context-free grammar} (CFG).
Context-free grammars (hereon simply \emph{grammars}) are a well-studied, widely used, and widely taught formalism for specifying the syntax of structured text.
Numerous automated techniques are available for turning grammars into input \emph{parsers} that decompose inputs according to the grammar, as well as turning grammars into \emph{producers} that generate conformant inputs.
All these make grammars an appealing choice for several tasks in computer science and software engineering.

As an example, \cref{fig:csv-snippet} shows a grammar for CSV (comma-separated values\footnote{We directly use the grammar from \fandango{}~\cite{zamudio2025fandango}, where the authors chose to use a CSV format separated by semicolons.}) files, a common data format.
This grammar can be easily used for parsing, but can it also produce valid CSV files?
Actually, no: the grammar does not ensure that all records (lines) have the same number of fields, thus na\"ively producing from this grammar will yield invalid CSV files.

\begin{figure}
    \centering
    \begin{minipage}{0.6\textwidth}
    \begin{densegrammar}
        <csv\_record> ::= <csv\_string\_list> `\textbackslash n'

        <csv\_string\_list> ::= <raw\_field> | <raw\_field> `;' <csv\_string\_list>
    \end{densegrammar}
    \end{minipage}
    \caption{A snippet from the CSV grammar for \fandango{}.}
    \label{fig:csv-snippet}
\end{figure}

Indeed, the downside of grammars is that they specify \emph{syntax} only.
Yet, many programming languages and data formats have intricate \emph{semantics} that cannot be captured by syntax alone either ergonomically or at all---in our case, the property that all CSV records must have the same number of fields.
While disregarding semantics is acceptable for parsing inputs, as one can simply check the semantics afterward, it is problematic for producing inputs, as many syntactically valid inputs may be semantically invalid.
Typical examples for such properties include
\begin{enumerate*}[label=(\arabic*)]
    \item XML documents, where the identifiers of opening tags must match closing tags;
    \item basic properties of programming languages, such as identifiers having to be declared before their use; or
    \item binary data formats, which frequently come with checksums and length encodings.
\end{enumerate*}
Consequently, grammars alone are insufficient to produce valid inputs for such languages.

A recent approach to test input generation, \emph{language-based testing}~\cite{dominicACM} (LBT), addresses the problem of semantic validity by combining grammars with semantic constraints---predicates over grammar elements that express input validity.
In the CSV example, such a constraint could express that all records have the same number of fields.
\cref{fig:csv-constraint} shows such a constraint in the constraint language of \isla, the first language-based input generator, effectively ensuring semantic validity.
In addition to guaranteeing semantic validity, constraints may also be used for specifying desirable generation properties.
For instance, \cref{fig:extra-csv-constraint} shows two constraints that force exactly three columns and three rows, respectively; testers can use such constraints to shape test inputs towards the desired goal.

\begin{figure}
    \begin{align*}
        &\forall r_1 \in \nonterm{csv\_record} \\
        &\quad{} \forall r_2 \in \nonterm{csv\_record} \\
        &\qquad \lvert r_1.\nonterm{csv\_string\_list}..\nonterm{raw\_field} \rvert = \lvert r_2.\nonterm{csv\_string\_list}..\nonterm{raw\_field} \rvert
    \end{align*}
    \caption{An LBT-style constraint over CSV, from the artifact of the original paper~\cite{zamudio2025fandango}. The cardinality operator measures the number of nodes matching the selected expression. One may read this as ``the number of \nonterm{raw\_field} within the recursive expansions of the \nonterm{csv\_string\_list} of $r_1$ is equal to that of $r_2$'s.''}
    \label{fig:csv-constraint}
\end{figure}

\begin{figure}
    \begin{minipage}{0.4\textwidth}
    \begin{align*}
        &\forall r \in \nonterm{csv\_record} \\
        &\quad{} \lvert r..\nonterm{raw\_field} \rvert == 3
    \end{align*}
    \end{minipage}
    \quad{}
    \begin{minipage}{0.4\textwidth}
    \begin{align*}
        &\forall f \in \nonterm{csv\_file} \\
        &\quad{} \lvert f..\nonterm{csv\_record} \rvert == 3
    \end{align*}
    \end{minipage}
    \caption{A pair of constraints which force exactly three columns and three rows, respectively.}
    \label{fig:extra-csv-constraint}
\end{figure}

However, being able to express constraints does not mean that we can solve them (i.e., produce inputs that satisfy them).
\isla used the SMT-LIB language to express constraints, and consequently, a symbolic SMT solver to solve constraints.
However, SMT solvers are known to be inefficient for string constraints~\cite{steinhofel2022input}, and \isla was found to be impractically slow for complex inputs.

To address this limitation, the \fandango{}~\cite{zamudio2025fandango} tool introduced a constraint language based on Python, and used evolutionary algorithms to solve constraints.
Instead of directly solving, inputs are generated and then iteratively mutated until they satisfy the constraints.
As this \emph{evolutionary language-based testing} (ELBT) removed the need for symbolic evaluation, it showed substantial performance improvements, with \fandango{} being faster than \isla{} by three orders of magnitude on common grammars.
But is \fandango{}'s ELBT strategy truly sufficient for testing complex systems in practice?

When we applied \fandango{} on the CSV grammar and the constraint from \cref{fig:csv-constraint}, we find that only inputs with one column or one row are produced.
These are, of course, valid according to the constraint, but are rather trivial satisfactions; obviously, \fandango{}'s evolutionary algorithm is inherently biased to produce the simplest inputs that satisfy the constraints.
As soon as we add the constraints from \cref{fig:extra-csv-constraint}, however, \fandango{} struggles to find valid inputs, taking over ten seconds to generate a single valid input in our experimental setup (see \cref{sec:experimental-setup}).
As the number and complexity of constraints grow, \fandango{}'s performance quickly degrades to minutes or hours per valid input (if it ever solves it at all), making it impractical for real-world testing with complex input languages.

The good news is that such performance issues can be addressed, elevating the performance of ELBT.
Our two key ideas to improve performance are:
\begin{enumerate}
    \item to compile grammar definitions into Rust, enabling the compiler to produce highly efficient implementations of operations generic over grammars; and
    \item to adopt better specialized evolutionary algorithms, which allow for the solving of higher-complexity constraint systems.
\end{enumerate}
These ideas make a difference.
Our \tool prototype produces complex inputs where \fandango{} fails, and produces dozens of complex, diverse, and valid inputs in seconds rather than hours, opening up language-based testing (and specification-based input generation in general) to input domains with complex semantics.

In detail, our contributions are as follows:

\begin{description}
    \item[Compiling grammars into high-performance code.] We introduce a compiler from context-free grammars to Rust types and implement operations on them generically.
    Our transformation
    \begin{enumerate*}[label=(\arabic*)]
        \item is complete and near-maximally optimal,
        \item absolutely enforces memory, type, and data correctness,
        \item exposes information to IDEs that enable development with generated types, and
        \item is easily extended to simplify generic and grammar-specific operations and fitness functions.
    \end{enumerate*}
    \item[Evolutionary algorithms for complex constraints.] We adopt a \emph{multi-objective} evolutionary algorithm to find optimal solutions, even when constraints conflict or differ in solving complexity, and allow for user-specified fitness functions.
    \item[A detailed evaluation.] We ablate our approach by profiling specific operations, improving performance by \fandangogenfactor{} over \fandango{} by way of our transformations.
    \item[A C case study.]
    Finally, we present a case study applying \tool{} on a C subset, demonstrating the impact of choosing appropriate evolutionary algorithms.
    In this, we demonstrate that our optimized, multi-objective form of \tool{} generates \multiValidAndSizedCompileAvgRs{} diverse, complex, and valid test inputs for a C compiler per minute, whereas neither single-objective \tool{} nor \fandango{} can produce any whatsoever.
\end{description}

The remainder of this paper is structured as follows:
\cref{sec:background} introduces the background necessary for the topics discussed;
\cref{sec:design} provides an overview of our transformation and optimization steps;
\cref{sec:profiling} evaluates \tool by profiling, ablation, and a direct analysis study for structure-specific optimizations;
\cref{sec:case-study} shows our case study, producing test cases for a C compiler.
After reflecting on threats to validity (\cref{sec:threats-to-validity}), limitations (\cref{sec:limitations}), and how our work relates to existing literature (\cref{sec:related-work}), \cref{sec:conclusion} closes with a conclusion and future work.

\section{Background} \label{sec:background}

Before getting into the proverbial meat and potatoes, several topics require at least a brief review to put the paper in context.
This section begins by pinpointing the role of Evolutionary Language-Based Testing (ELBT) in search-based testing and highlights why it is specifically desirable but understudied.
Then, we take a brief detour into evolutionary algorithms and their applications, especially highlighting problems involving structural inputs.
Finally, we provide a basic overview of the Rust programming language and the key language features that are leveraged over the course of the paper.

\subsection{Fuzz Testing and Evolutionary Language-Based Testing}

Fuzz testing (or \textit{fuzzing}) is an automated software testing technique that aims to uncover defects by generating large volumes of inputs and monitoring program behavior for crashes or anomalies.
The general idea of fuzzing is that one may search the program space of a tested program by progressively searching over its inputs.
While fuzzing has uncovered thousands of bugs across a wide range of systems, the vastness of most input spaces and the sparsity of defect-triggering inputs make effective exploration a major challenge.

Traditional fuzzers like \AFL~\cite{aflfuzz} and \libfuzzer~\cite{libfuzzer} mutate inputs lexically, evolving random mutations of seed inputs through a search strategy similar to novelty-guided search~\cite{novelty-search,novelty-fuzzing}.
Although these approaches are efficient, they rely heavily on high-quality seeds and often struggle with structured input formats or semantic correctness.

\subsubsection{Grammar-Based Fuzzing}

To address these limitations, \textit{grammar-based fuzzing} uses context-free grammars (CFGs) to guide input generation, ensuring that generated inputs are syntactically valid.
CFGs, formalized by Chomsky~\cite{Chomsky_1956}, describe input formats using production rules over terminals and nonterminals.
For instance, \cref{fig:expr-grammar} shows a grammar for generating simple arithmetic expressions.
Grammar-based fuzzing is fast and reliable, capable of producing valid inputs in large quantities, especially useful when fuzzing programs that parse complex structured inputs~\cite{burkhardt1967,purdom1972,godefroid2008grammar, wang2019superion, hodovan2018grammarinator, aschermann2019nautilus}.

Moreover, grammar-based generation provides control over the production of inputs.
This allows fuzzers to reason about and track exploration of the input space using metrics like \emph{grammar coverage}, which quantifies the diversity of derivation trees exercised during fuzzing. 
The common $k$-path metric~\cite{havrikov2019systematically} quantifies the coverage of a particular grammar by measuring the number of observed paths of length $k$ or less across a forest of derivation trees---in this case, our inputs.
This allows us to measure the diversity of the inputs we have already produced and guide future generations towards paths we have not already explored.

\subsubsection{Fast Grammar Fuzzing}

One advantage of grammar-based fuzzing is the speed at which it can generate syntactically valid inputs.
A naïve implementation of a grammar-based fuzzer would simply have a working string and continuously replace nonterminals with randomly chosen productions until no nonterminals remain.
However, this approach is inefficient, as it requires numerous string replacements and concatenations.

A much more efficient approach is to compile the grammar into \emph{producer functions} that can generate inputs directly.
In this approach, each production rule is turned into a function that first randomly chooses one of the alternatives.
Then, for terminals, it produces the corresponding string; for nonterminals, it calls the corresponding producer function.
Further compiler optimizations, such as inlining, constant folding, and stack elimination, result in efficient producers that can generate millions of syntactically valid inputs per second, as exhibited by the F1~\cite{Gopinath_Zeller_2019} and F0~\cite{Falk_2025} fuzzers.

\subsubsection{Semantic Fuzzing}

Despite their speed, grammar-based fuzzers have important limitations that make them unsuitable for many real-world applications.
Most importantly, the expressiveness of context-free grammars is limited to \emph{syntax}, meaning that they can only describe the structure of inputs, but not \emph{semantic features} such as checksums, numeric properties, or cross-references.

It is here that we return to what was already discussed in the introduction.
Language-based testing (LBT) strategies address the aforementioned limitations by augmenting grammars with \emph{semantic constraints}~\cite{dominicACM}---logic over grammar elements that must be satisfied for an input to be semantically valid.
\isla{}~\cite{steinhofel2022input} solves such constraints using the \Zthree{} SMT solver~\cite{demoura2008z3}; \fandango{}~\cite{zamudio2025fandango} uses evolutionary algorithms to find solutions.
Both of these are slower than fast grammar-based fuzzers, but then again, they produce outputs that are valid syntactically and semantically.

\subsection{Evolutionary Algorithms}

Evolutionary algorithms are a class of meta-heuristic algorithms~\cite{heuristic} which attempt to solve optimization problems by ``evolving'' a population of candidate solutions.
Numerous evolutionary algorithms perform a variety of optimization tasks.
In this paper, we primarily use and discuss \emph{genetic algorithms} (GA).
In GA, the actual evolution procedure varies according to the problem being solved, but in the context of this paper, there are two primary operators: mutation, in which a given population member is independently modified by regenerating input components, and crossover, where one to many candidates are produced by exchanging components of members of the previous population.
At each step of the evolution, the population is then reduced to a locally optimal set of solutions from the generated candidates.
The evolution is then progressively stepped until a halting condition is reached (typically, by objective(s) reaching a local optimum or an execution limit).

One chooses different evolutionary algorithms according to the class of optimization problem being solved.
In this paper, we deal with two: single- and multi-objective optimization.
The definitions for these are somewhat obvious; single-objective has only one objective, multi-objective has many.
Yet, how we actually accomplish optimization for these problems is quite different.
For a single objective, there is only one variable being optimized, and thus there is a total ordering whereby one may select ``best'' candidates for evolution.

For multi-objective, there is often no total ordering.
In \tool{}, we use algorithms which select candidates by Pareto optimality~\cite{pareto-front-definition}.
When we say that a candidate solution is Pareto optimal, we mean that there is no other solution that dominates (i.e., is strictly better than\footnote{``Better'' here is objective-dependent. Our EA implementations allow users to specify heterogeneously-typed objectives, and thus, it is specific per-objective as to what ``better'' means.}) this solution.
We can organize candidate solutions into so-called Pareto \emph{fronts}, sets of candidate solutions where there exists no other candidates that dominate any member of that set.
A population of solutions, then, may be organized into an ordered list of Pareto fronts by descending domination order, a process referred to as \emph{non-dominated sorting}~\cite{nsga2}.
We emphasize again that this is not the only method for multi-objective optimization in EA, but it is what we utilize.

\subsection{The Rust Programming Language}

Rust~\cite{rustbook} has seen increasing popularity in recent years in hobbyist, industrial, and academic applications due to its rich type system, a strong memory model which eliminates memory safety issues, and strong optimization capabilities~\cite{stackover}.
This paper deals heavily with program generation aspects, specifically within the Rust language.
\ifarxiv
While we provide a brief overview here, code snippets provided in the paper may require greater knowledge of the language than what could be provided in a brief background.
We refer those who wish to have a deeper understanding of Rust's language features to the Rust Book~\cite{rustbook} and provide a high-level, if slightly simplified, overview of the aspects of the language most relevant to our work below.

\subsubsection{Structs, Enums, Traits, and Generics}

Regarding the type system of Rust, we discuss in particular these four primitives.

\begin{description}
    \item[Structs] are structured types with zero to many fields, mostly analogous to (concrete) classes from object-oriented programming languages~\cite{rentsch1982object}, but without inheritance.
    \item[Enums] are types that represent a selection over structured data; different \emph{variants} of enums contain different structured data.
    This is not analogous to enums from most popular languages, like C or Java, as Rust enums do not merely represent scalar values but instead represent the \emph{tagged union} concept from functional languages~\cite{russell2001events}.
    \item[Traits] are not types, but rather a concept which expresses functionality that a type can have and (meta-)data associated with the type itself (i.e., not specific to instances).
    Traits are \emph{implemented} for specific types to imbue them with functionality and associated data.
    Unlike similar paradigms in object-oriented languages (e.g., Java interfaces), these are applied post-facto and can be applied to types not present within the current Rust software unit.
    Because traits are applied to types, we may also implement traits for many types simultaneously via \emph{generics}.
    \item[Generics] allows for the definition of generic behavior by specifying a named ``stand-in'' for a concrete type, often with constraints which specify required traits for a given concretization of that behavior.
    This allows for the definition of types, traits, methods, and trait implementations that have specified behavior for all types (including those downstream) that satisfy the constraints specified on those generics.
\end{description}

\subsubsection{The Ownership Model}

A concept unique to Rust among mainstream programming languages is the \emph{ownership model}, which is a memory management model that assigns ownership of variables to particular scopes.
A variable is \emph{owned} if a scope is responsible for the destruction of that variable and ownership may be transferred to other scopes (e.g., by passing to or returning from a function).
\emph{Borrowing} is the act of maintaining a reference to a variable, and can be done via \emph{immutable} or \emph{mutable} borrows, which disallow and allow writing to the variable, respectively.
When a variable is borrowed, its ownership may not be transferred, and it may not be destroyed (as it may move in memory or invalidate the existing reference).
Similarly, when a variable is immutably borrowed, it may only be immutably borrowed again, and when mutably borrowed, it may not be borrowed---mutably or no---again.
These rules apply both to instances as a whole and to fields of those instances, meaning that one can, e.g., independently mutably borrow fields of an instance, and borrowing/ownership transfer of the instance itself is forbidden to prevent the invalidation of these borrows.
In complex situations where borrows of variables persist within other types (e.g., a struct which has a field for which the type is a reference of a type), the reference is annotated with a \emph{lifetime} which symbolically represents its scope.

All of these properties allow for the static enforcement of borrow semantics (i.e., you cannot compile a Rust program that violates these rules).
Encoding access to variables based on borrow semantics is the idiomatic fashion by which one may statically enforce desired program properties of data management.
For example, memory and resources are managed through this system (vis-\`a-vis the RAII model~\cite{RAII}) via the destruction of the variable upon its exiting of the owning scope (when it is ``dropped''), and ensure that memory and resources remain available while used.

\subsubsection{Compiler Optimizations}

\fi

We carefully orient our design around the optimizations of Rust to achieve extreme performance while retaining the ability to be very expressive in the code we develop.
The most critical of the optimizations made available to us is \emph{monomorphization}~\cite{monomorphization}.
In many programming languages that support generics, generic definitions are concretized once and may be interacted with using different types.
When these types are used within the generic definition, access is first dynamically resolved (e.g., via virtual pointer table lookup~\cite{dispatcher}) before being utilized.
In Rust, generics undergo monomorphization, wherein each statically reachable variant of that definition has its own concrete implementation.
This allows the compiler to optimize aggressively and simplify code according to the individual functionality present in these concrete forms (\emph{instantiations}) of the generic definitions.

Other optimizations also severely affect the performance of our input generation routines, but we observe this in this paper primarily in tandem with monomorphization.
Bounds inference and other classical program analysis strategies common to Rust can identify where checks are not needed and simply remove them, and similarly allow for the merging of operations with compile-time inferrable effects.
Rust employs many such optimizations, but can only do so when there is sufficient context in the source code to apply them.
Our design below attempts to maximize optimizability by encoding as much information as possible into each generic function, allowing the compiler to infer optimizations from the structure of the grammar itself.

\section{Design} \label{sec:design}

As the stated goal of our paper, we want to make our strategy as fast as possible in both producing input components and solving constraints.
At the same time, we do not want to prevent the use of our strategy because the generated code is too complicated or too far removed from the original structure.
This is the central theme of this section: optimizing both for performance and usability, and it begins at the core of our approach.

\subsection{A Source-to-Source Transformation from Context-Free Grammars to Rust Types} \label{sec:grammar-to-rust}

As hinted in the introduction, our design centers around generating Rust types that we may then reuse to achieve arbitrary testing goals.
Indeed, we design our transformation such that grammar-specific knowledge encoded into the type system is used to optimize operations taken on user-provided grammars.
As a bonus, a correct transformation from CFG to Rust types will, in effect, ensure an input's syntactic correctness by its type correctness, reducing developer error.

For explaining this transformation, we will forego a formal definition of CFGs and instead focus on the \emph{structure} of such grammars.
Classically, a nonterminal may be defined by (1) a nonterminal (referring to another nonterminal definition), (2) a terminal, or (3) a concatenation or (4) an alternation (i.e. ``pick exactly one of these'') over other primitives.
In many modern tools in which context-free grammars are defined, including \fandango{} and \isla{}, we find repetition primitives as well: (5) a Kleene star operator, for zero to many repetitions of another primitive; (6) a plus operator for one to many; (7) a range operator, expressing a repetition of $[n_1,n_2)$ primitives; and (8) an option operator, expressing zero or one.
While all repetitions could be ``canonicalized'' into the first four primitives, repetitions allow for more ergonomic expression of sequences.
As they are quite useful in practice, we include these primitives as well.

In Rust, we find the corresponding equivalents:

\begin{multicols}{2}
\begin{enumerate}
    \item A struct with a single field,
    \item A zero-sized type (ZST) with an associated constant representing the literal,
    \item A struct with one field per member of the concatenation,
    \item An enum with one variant per member of the alternation,
    \itemrange{2} A \texttt{Vec} (i.e. an array) of repeated elements,
    \item An \texttt{Option} of a single element.
\end{enumerate}
\end{multicols}

Because every primitive for defining a nonterminal is represented, we can be confident that applying this transformation will be able to represent arbitrary grammars.
In practice, this is accomplished by first parsing a grammar into a representative graph before transforming the graph into the Rust types.

\subsubsection{Grammar to Graph Transformation} \label{sec:grammar-to-graph}

Transforming grammars into graphs is not conceptually new, but there are some important steps that take place along the way, which give motivation for later steps.
To demonstrate the transformation process, we define our own grammar, \emph{expr}, in \cref{fig:expr-grammar,} which we will progressively convert.

To actually perform the conversion, each nonterminal is first expanded into a subgraph which represents its local expansion, where a node's weight represents the operation or (non)terminal and its outgoing edges are weighted by the index of the expansion (see \cref{fig:expr-nonterm-graph}).
These graphs are then merged at the nonterminals, forming a larger graph that represents the whole grammar (see \cref{fig:expr-graph}).

\begin{figure}
    \begin{minipage}{0.5\textwidth}
    \begin{densegrammar}
    <start> ::= <expr>
    
    <expr> ::= <number> `+' <expr> | <number>
    
    <number> ::= `0' | <non_zero><digit>*
    
    <non\_zero> ::= `1' | `2' | `3' | `4' | `5' | `6' | `7' | `8' | `9'
                 
    <digit> ::= `0' | <non_zero>
    \end{densegrammar}
    \caption{Definition of the \emph{expr} grammar.}
    \label{fig:expr-grammar}
    \end{minipage}
    \hfill
    \begin{minipage}{0.45\textwidth}
    \centering
    \tiny
    \begin{tikzpicture}[on grid,auto]
        \node[state] (expr) {\nonterm{expr}};
        \node[state] (expr-alt) [right=of expr] {$|$};
        \node[state] (expr-concat-0) [above=of expr-alt] {$\sim$};
        \node[state] (number) [right=1.25cm of expr-alt] {\nonterm{number}};
        \node[state] (plus) [right=of expr-concat-0] {``$+$''};
        
        \path[->]
            (expr) edge node [below] {0} (expr-alt)
            (expr-alt) edge node [right] {0} (expr-concat-0)
            (expr-alt) edge node [below] {1} (number)
            (expr-concat-0) edge node [above left] {2} (expr)
            (expr-concat-0) edge node [above right] {0} (number)
            (expr-concat-0) edge node [above] {1} (plus)
            ;
    \end{tikzpicture}
    \vspace{1.3em}
    \caption{Graph representing \nonterm{expr}.}
    \label{fig:expr-nonterm-graph}
    \end{minipage}
\end{figure}

\begin{figure}
    \centering
    \tiny
    \begin{tikzpicture}[on grid,auto]
    \node[state] (start) {\nonterm{start}};
    \node[state] (expr) [right=1.25cm of start] {\nonterm{expr}};
    \node[state] (expr-alt) [right=of expr] {$|$};
    \node[state] (expr-concat-0) [above=of expr-alt] {$\sim$};
    \node[state] (number) [right=1.25cm of expr-alt] {\nonterm{number}};
    \node[state] (plus) [right=of expr-concat-0] {``$+$''};
    \node[state] (number-alt) [right=1.25cm of number] {$|$};
    \node[state] (number-zero) [above=0.75cm of number-alt] {``$0$''};
    \node[state] (number-concat-1) [right=of number-alt] {$\sim$};
    \node[state] (number-kleene) [above=0.75cm of number-concat-1] {$*$};
    \node[state] (digit) [above=of number-kleene] {\nonterm{digit}};
    \node[state] (digit-alt) [right=1.25cm of digit] {$|$};
    \node[state] (digit-zero) [right=0.75cm of digit-alt] {``$0$''};
    \node[state] (nonzero) [right=1.25cm of number-concat-1] {\nonterm{nonzero}};
    \node[state] (nonzero-alt) [right=4.5cm of nonzero] {$|$};
    \node[state] (nonzero-5) [above=1.75cm of nonzero-alt] {``$5$''};
    \node[state] (nonzero-4) [left=0.75cm of nonzero-5] {``$4$''};
    \node[state] (nonzero-3) [left=0.75cm of nonzero-4] {``$3$''};
    \node[state] (nonzero-2) [left=0.75cm of nonzero-3] {``$2$''};
    \node[state] (nonzero-1) [left=0.75cm of nonzero-2] {``$1$''};
    \node[state] (nonzero-6) [right=0.75cm of nonzero-5] {$\dots$};
    
    \path[->,dotted]
        (expr-concat-0) edge node [above left] {2} (expr)
        (number-kleene) edge node [right] {0..} (digit)
        ;
    \path[->]
        (start) edge node [below] {0} (expr)
        (expr) edge node [below] {0} (expr-alt)
        (expr-alt) edge node [right] {0} (expr-concat-0)
        (expr-alt) edge node [below] {1} (number)
        (expr-concat-0) edge node [above right] {0} (number)
        (expr-concat-0) edge node [above] {1} (plus)
        (number) edge node [below] {0} (number-alt)
        (number-alt) edge node [right] {0} (number-zero)
        (number-alt) edge node [below] {1} (number-concat-1)
        (number-concat-1) edge node [below] {0} (nonzero)
        (number-concat-1) edge node [right] {1} (number-kleene)
        (digit) edge node [above] {0} (digit-alt)
        (digit-alt) edge node [above] {0} (digit-zero)
        (digit-alt) edge node [right] {1} (nonzero)
        (nonzero) edge node [below] {0} (nonzero-alt)
        (nonzero-alt) edge[in=270,out=135.0] node [below] {0} (nonzero-1)
        (nonzero-alt) edge[in=270,out=123.75] (nonzero-2)
        (nonzero-alt) edge[in=270,out=112.5] (nonzero-3)
        (nonzero-alt) edge[in=270,out=101.25] (nonzero-4)
        (nonzero-alt) edge[in=270,out=90.0] (nonzero-5)
        (nonzero-alt) edge[in=270,out=78.75] node [below right] {5--8} (nonzero-6)
        ;
    \end{tikzpicture}
    \caption{Graph formed by merging nonterminal graphs like that of \cref{fig:expr-nonterm-graph}. Indirect edges are dotted.}
    \label{fig:expr-graph}
\end{figure}

\subsubsection{Combined Graph to Rust Transformation} \label{sec:graph-to-rust} \label{sec:indirection}

Though naïvely transforming grammars into Rust types is tempting, there is a fatal flaw: memory layout.
When determining the size of a type in memory, Rust considers its largest possible manifestation to support arbitrary modifications of the value in-place.
As a result, any fields that would cause a cycle (and therefore an infinitely-sized type) must be placed within a memory allocation, reducing the size of the field within the parent structure to a pointer.

Indirection requires extra operations for performing (de)allocation and slows traversal by accessing non-contiguous memory likely outside of the cache.
Optimally, we would perform as few allocations as possible.
Preemptively removing the edges where indirection is forced (namely, at repetitions), we compute the feedback arc set~\cite{kudelic2022feedback} of the graph, which reveals the smallest set of edges that need to be removed to ensure there are no cycles in the graph.
By introducing indirection at these edges, we remove all cycles in our type structure, which guarantees that all produced types will be of finite size with the fewest number of indirection sites.
This is represented once again in \cref{fig:expr-graph} by the dotted edges.
Otherwise, we follow the steps outlined at the beginning of \cref{sec:grammar-to-graph} to complete the transformation.
To avoid invalidating existing code when a user changes a grammar, we introduce accessor patterns that prevent the user from needing to reason about the indirection.

With these steps completed, we now have types for which instances represent derivation subtrees.
Instantiating a node is now exactly equivalent to instantiating a derivation tree in the original grammar.

\subsubsection{Opaque Types}

In some situations, it is necessary to refer to instances of our generated types without referring to their concrete type.
For example, suppose that one wishes to perform level-order traversal of a derivation tree\footnote{Example code for this operation is provided in the artifact.}.
Part of this process is to build a work queue of subtrees yet to be traversed, but heterogeneously-typed lists cannot be constructed dynamically in Rust.
Moreover, as we show later in \cref{sec:visitors}, there are situations where one wishes to implement a generic over a whole grammar rather than on specific types.
To address both of these, we introduce two grammar-specific opaque types, implemented as enums, which respectively wrap either an immutable or a mutable reference to each explicit node type.
Now, we can produce lists of heterogeneously-typed lists over subtrees of a single derivation tree, with exclusive mutability guarantees from traversal implementation, or implement traits generically over the whole grammar via the corresponding opaque type(s).
We associate these opaque types with nodes over generated trait implementations and provide helper functions to up- and down-``cast'' nodes accordingly.

\ifarxiv
\subsubsection{Unevaluated Benefits of Direct Transformation}

By construction, our transformation integrates quite fluidly with IDEs and helpful compiler hints from Rust.
Users can often observe typing hints within the IDE that directly inform the user of precisely what type is being manipulated.
In addition, we generate and attach documentation to the generated source code that allows the user to directly inspect the region in the grammar that corresponds to each type.
As a result, traversing and manipulating typed nodes concretely is very straightforward and well-documented, and the compiler gives useful feedback whenever the user makes mistakes while handling concrete nodes.
Though not formally evaluated in this paper, this integration massively simplified and sped up our grammar-specific usage in \cref{sec:profiling,sec:case-study}, reducing our debugging time to effectively zero.
Moreover, because our transformations are standalone, they (and all described operations implemented in the rest of \cref{sec:design}) are executable on \emph{any supported Rust target}!
We were able to use \tool{} on several architectures and operating systems, and we include an example of embedding it as firmware in the artifact.
\fi

\subsection{Samplers and Generators} \label{sec:generation}

Users may now directly instantiate derivation trees by using our generated constructors and/or accessors.
By construction, any type instantiation necessarily reflects a syntactically valid derivation tree in the original grammar from the type's corresponding node.
In LBT, we often want to generate derivation trees, and having to write code to do so for each type is unwieldy.
To avoid this, we define two additional patterns, \emph{samplers} and \emph{generators}.
By implementing their own samplers and generators, users can influence how nodes are generated without having to explicitly construct them.
Our design ensures that they may do so with minimal performance impact.

Since the structure of the nodes is known in advance, node generation can be considered merely as a sequence of choices about which form of a given node to produce~\cite{enumerating-cfgs}.
In our context-free grammars, there are two choices: (1) at alternations, which variant to expand, and (2) at repetitions, how many elements to produce.
\emph{Samplers} (which can be as simple as a random number generator) inform generation as to which choices to make, and users may use the default generation strategy in combination with their custom sampler implementations to guide this process to their will.
In our generation model, each node has a pre-defined implementation for generation based on queries to the provided sampler.
When users want to customize the generation process itself (e.g., by constructing a node directly or tampering with sampling for a particular derivation tree), they may specify custom \emph{generators}.
When the user specifies a list of generators, each generator is queried to check if it has a custom generation pattern for the currently generated node.
If one does, it returns the generated node; otherwise, the default strategy recurses using the sampler to determine structure and attempts to generate the subsequent nodes with the same list of generators again.
In short: the samplers select which choice is taken, and the generators determine what choices are available and how that choice is applied.
Users may either rely on our default sampler and generator implementations, or provide their own to accomplish particular generation objectives; we provide several examples in the artifact.

\ifarxiv
\subsubsection{Example: Remediating Unwanted Structural Generation Bias}

To elucidate how samplers and generators work together in practice, we describe the \emph{flattener} generation pattern.
Consider the \nonterm{digit} definition from \cref{fig:expr-grammar}: a \nonterm{digit} is defined as either a ``$0$'' literal or a \nonterm{non\_zero}.
If the sampler is unaware of the grammar, it will first decide between the ``$0$'' and \nonterm{non\_zero} with a $\nicefrac{1}{2}$ chance of each; if \nonterm{non\_zero}, each subsequent terminal will be selected with a $\nicefrac{1}{9}$ chance.
Thus, $\nicefrac{1}{2}$ of \nonterm{digit} expansions will be `$0$' and $\nicefrac{1}{18}$ each other digit literal.

To resolve this, we implement a flattener generator which ``flattens'' a tree of alternations such that each of its descendant terminals is equally likely.
Since flattening is an operation that is inferable from structure, the flattener generator is generic and, when instantiated over a specific node, can precompute a table of sequences necessary to reach all subsequent terminals uniformly.
Computed sequences (actually integerized stacks~\cite{enumerating-cfgs}) for \nonterm{digit} are shown alongside terminals in \cref{fig:digit-flattened} and the actual generation process with sampling steps is portrayed in \cref{fig:digit-generation}.

\begin{figure}[H]
    \begin{minipage}{0.34\textwidth}
    \centering
    \tiny
    \begin{tikzpicture}[on grid,auto]
        \node[state] (digit) {\nonterm{digit}};
        \node[state] (digit-alt) [right=of digit] {$|$};
        \node[state] (nonzero-5) [right=of digit-alt] {``$5$''};
        \node[state] (nonzero-4) [below=0.675cm of nonzero-5] {``$4$''};
        \node[state] (nonzero-3) [below=0.675cm of nonzero-4] {``$3$''};
        \node[state] (nonzero-2) [below=0.675cm of nonzero-3] {``$2$''};
        \node[state] (nonzero-1) [below=0.675cm of nonzero-2] {``$1$''};
        \node[state] (digit-zero) [below=0.675cm of nonzero-1] {``$0$''};
        \node[state] (nonzero-6) [above=0.675cm of nonzero-5] {``$6$''};
        \node[state] (nonzero-7) [above=0.675cm of nonzero-6] {``$7$''};
        \node[state] (nonzero-8) [above=0.675cm of nonzero-7] {``$8$''};
        \node[state] (nonzero-9) [above=0.675cm of nonzero-8] {``$9$''};

        \node [right=0.3cm of digit-zero,anchor=west] {\texttt{stack[0] = seq[0]}};
        \node [right=0.3cm of nonzero-1,anchor=west] {\texttt{stack[1] = seq[1,0]}};
        \node [right=0.3cm of nonzero-2,anchor=west] {\texttt{stack[2] = seq[1,1]}};
        \node [right=0.3cm of nonzero-3,anchor=west] {\texttt{stack[3] = seq[1,2]}};
        \node [right=0.3cm of nonzero-4,anchor=west] {\texttt{stack[4] = seq[1,3]}};
        \node [right=0.3cm of nonzero-5,anchor=west] {\texttt{stack[5] = seq[1,4]}};
        \node [right=0.3cm of nonzero-6,anchor=west] {\texttt{stack[6] = seq[1,5]}};
        \node [right=0.3cm of nonzero-7,anchor=west] {\texttt{stack[7] = seq[1,6]}};
        \node [right=0.3cm of nonzero-8,anchor=west] {\texttt{stack[8] = seq[1,7]}};
        \node [right=0.3cm of nonzero-9,anchor=west] {\texttt{stack[9] = seq[1,8]}};
    
        \path[->]
            (digit) edge node [above] {0} (digit-alt)
            (digit-alt) edge[out=225,in=180] node [left] {0} (digit-zero)
            ;
        \path[->,dashed]
            (digit-alt) edge[in=180,out=270] node [left] {1} (nonzero-1)
            (digit-alt) edge[in=180,out=292.5] (nonzero-2)
            (digit-alt) edge[in=180,out=315.0] (nonzero-3)
            (digit-alt) edge[in=180,out=337.5] (nonzero-4)
            (digit-alt) edge[in=180,out=0] (nonzero-5)
            (digit-alt) edge[in=180,out=22.5] (nonzero-6)
            (digit-alt) edge[in=180,out=45.0] (nonzero-7)
            (digit-alt) edge[in=180,out=67.5] (nonzero-8)
            (digit-alt) edge[in=180,out=90.0] node [left] {9} (nonzero-9)
            ;
    \end{tikzpicture}
    \caption{A ``flattened'' \nonterm{digit}.}
    \label{fig:digit-flattened}
    \end{minipage}
    \hfill
    \begin{minipage}{0.65\textwidth}
    \centering
    \tiny
    \tikzstyle{generator} = [rectangle, minimum width=1.5cm, minimum height=0.5cm, text centered, draw=black, fill=white]
    \tikzstyle{sampler} = [rectangle, minimum width=1cm, minimum height=0.5cm, text centered, draw=black, dotted, fill=white]

    \begin{tikzpicture}[on grid,auto]
        \node (entry) {};
        \node[state] (digit) [below=of entry] {\nonterm{digit}};
        \node[state] (digit-alt) [below=1.5cm of digit] {$|$};
        \node[state] (digit-zero) [right=of digit-alt] {``$0$''};
        \node[state] (nonzero) [below=of digit-alt] {\nonterm{nonzero}};
        \node[state] (digit-alt-2) [below=of nonzero] {$|$};
        
        \node[state] (nonzero-5) [below=of digit-alt-2] {$...$};
        \node[state] (nonzero-6) [right=0.675cm of nonzero-5] {``$6$''};
        \node[state] (nonzero-7) [right=0.675cm of nonzero-6] {``$7$''};
        \node[state] (nonzero-8) [right=0.675cm of nonzero-7] {``$8$''};
        \node[state] (nonzero-9) [right=0.675cm of nonzero-8] {``$9$''};

        \node[generator] (default) [left=1.5cm of digit] {\emph{default generator}};
        \node[generator] (flattener) [left=1.5cm of default] {Flattener\nonterm{digit}};

        \node[sampler] (sampler) [left=2cm of flattener] {\emph{sampler}};

        \node[generator] (default-2) [left=1.5cm of digit-alt] {\emph{default generator}};
        \node[sampler] (seq-1) [left=1.5cm of default-2] {\texttt{seq[1,6]}};

        \node[generator] (default-4) [left=1.5cm of nonzero] {\emph{default generator}};

        \node[generator] (default-3) [left=1.5cm of digit-alt-2] {\emph{default generator}};
        \node[sampler] (seq-2) [left=1.5cm of default-3] {\texttt{seq[6]}};

        \node (empty) [below=of seq-2] {\emph{exhausted flattened sampler}};

        \node[draw, rectangle callout, callout relative pointer={(0.2,-0.55)}, text width=3.4cm] (callout) [above=1.25cm of flattener] {Hey, I know how to generate this!\\This is an ``alternation'' of 10 choices.\\Keep generating, but use my sampler.};

        \path[->]
            (entry) edge node [right] {$...$} (digit)
            (digit) edge node [right] {0} (digit-alt)
            (digit-alt) edge node [right] {1} (nonzero)
            (nonzero) edge node [right] {0} (digit-alt-2)
            (digit-alt-2) edge[in=90,out=315.0] node [below] {6} (nonzero-7)
            ;
        \path[->,dotted]
            (digit-alt) edge (digit-zero)
            (digit-alt-2) edge[in=90,out=270.0] (nonzero-5)
            (digit-alt-2) edge[in=90,out=292.5] (nonzero-6)
            (digit-alt-2) edge[in=90,out=337.5] (nonzero-8)
            (digit-alt-2) edge[in=90,out=360.0] (nonzero-9)
            ;
        \path[->,dashed]
            (flattener) edge[bend right] node [above] {sample\_alt(10)} (sampler)
            (sampler) edge[bend right] node [below] {7} (flattener)
            (flattener) edge node [below left] {\texttt{stack[7] = seq[1,6]}} (seq-1)
            (default-2) edge[bend right] node [above] {sample\_alt(2)} (seq-1)
            (seq-1) edge[bend right] node [below] {\texttt{1}} (default-2)
            (seq-1) edge (seq-2)
            (default-3) edge[bend right] node [above] {sample\_alt(9)} (seq-2)
            (seq-2) edge[bend right] node [below] {\texttt{6}} (default-3)
            (seq-2) edge (empty)
            ;
    \end{tikzpicture}
    \vspace{0.8em}
    \caption{Process of generating a \emph{flattened} \nonterm{digit}.}
    \label{fig:digit-generation}
    \end{minipage}
\end{figure}

In \cref{fig:digit-generation}, generation enters and attempts to generate a \nonterm{digit}.
The flattener over \nonterm{digit} identifies that it can generate this node, then samples over its list of precomputed generation sequences and invokes the default generator with a sampler that returns the sampled sequence, eventually resolving the ``$7$'' nonterminal.
As reasoning about whether a node is generable by this flattener is trivially true for \nonterm{digit} and trivially false otherwise, the compiler can infer that it is only necessary to invoke the generator for \nonterm{digit}s.
Moreover, this demonstrates how samplers and generators may symbiotically influence each other.
If the initial sampler in \cref{fig:digit-generation} is tracking the covered variants of each alternation, the flattener effectively notifies it that \nonterm{digit} is treated as an alternation with 10 variants.
That sampler may then track that variant 7 of \nonterm{digit} has already been sampled to tune future samplings, according to the sampler provided by the user.
\fi

\subsection{Visitors} \label{sec:visitors}

Now that we have a set of types corresponding to a grammar with APIs with which to construct and generate them, we want to be able to use these types to accomplish some goal.
This means both being able to inspect and manipulate instances of these types, preferably with the ability to implement such operations generically when specific knowledge of the grammar is not needed.
A typical solution to this problem is to use a \emph{visitor pattern}~\cite{visitorpattern}, wherein a ``visitor'' type implements a callback for each type to be visited.
While theoretically we could generate visitor traits per-grammar, this would make it effectively impossible to create a generic visitor that can act on arbitrary grammars.

Instead, we define a visitor trait generic over a grammar's opaque type, which itself contains a function generic over an explicit type.
Rather than describe this pattern conceptually (by which we will miss many details), we present an example visitor in \cref{fig:write-visitor-impl}.
This figure shows the implementation of a visitor that serializes a derivation tree into its corresponding string.
Visitors that handle mutable nodes have a nearly identical structure.

\begin{figure}
    \centering
    \small
    \begin{cminted}{rust}
impl<W, T> Visitor<T> for WriteVisitor<W>
where
    T: VisitableChildren<T>, W: io::Write,
{
    type Continue = Self;    // allows iterative traversal
    type Break = Infallible; // will not ever break control flow
    type Error = io::Error;  // forwards errors from the write
    
    fn visit<'program, N>(mut self, node: &'program N, _: usize)
        -> VisitResult<Self, T>
    where
        N: Node<Type<'program> = T>,
        T: From<&'program N> + AsNodeRef<N>,
    {
        match node.definition() { // get the grammar entry
            FandangoNode::String(s) => { // terminal!
                // self.output is a W; s.inner() is a &[u8]
                self.output.write_all(s.inner())?;
                Ok(ControlFlow::Continue(self))
            }
            _ => { // something else
                node.opaque().visit_each(self)
            }
        }
    }
}

/// The result type returned by visitors.
pub type VisitResult<V, T>
where
    V: Visitor<T>,
= Result<ControlFlow<V::Break, V::Continue>, V::Error>;
    \end{cminted}
    \caption{An implementation of a visitor that writes the derivation tree to an output.}
    \label{fig:write-visitor-impl}
\end{figure}

This implementation may at first appear intimidating with its various generics, associated types, and lifetimes, but its design ensures the following qualities:

\begin{description}
    \item[Return type consistency.]
    Suppose we constrain our trait over \texttt{N} (the node type) and not \texttt{T} (its opaque type).
    The type system can no longer ensure that this visitor will have the same \texttt{Continue}, \texttt{Break}, or \texttt{Error} types returned from the \texttt{visit} function since the implementation of the trait may differ between specific \texttt{N}.
    As such, we force consistency across a single grammar by making the visitor generic over the opaque type, which is necessarily the same for all nodes of the same grammar.
    \item[Lifetime consistency.]
    Suppose we want to return a node from this visitor.
    By specifying that \texttt{N::Type<'program> = T} (i.e., that the immutable opaque type of \texttt{N} for lifetime \texttt{'program} is \texttt{T}), we effectively encode the lifetime \texttt{'program} into \texttt{T}.
    Thus, when an instance of an opaque type is returned (e.g., in \texttt{Self::Break}), the borrow checker may infer that the returned opaque value contains the lifetime \texttt{'program} (because it contains a \texttt{\&'program N}).
    \item[Constraint specification.]
    In Rust, one may not specify any further constraints on type arguments of generic functions (in this case, \texttt{N}).
    By making the visitor generic over \texttt{T}, we may constrain \texttt{T} at the implementation level and still constrain (and therefore access additional methods on) the opaque type.
    This allows, for example, the implementation of a generic visitor that does a task on any grammar with the required features (including user-defined features) as constrained over \texttt{T}.
    \item[Performance.]
    Since monomorphization of the \texttt{visit} function will take place over the generic type \texttt{N}, the emitted code will be optimized (potentially recursively) over \texttt{N}.
    As such, opaque type instances which trivially refer to \texttt{node} will be optimized away, allowing for e.g. zero-cost conditional casting from \texttt{N} to a user-specified explicit type, optimized traversal over children nodes, and so on.
    This is highlighted especially in \cref{sec:compiler-opts-inspection}.
\end{description}

Altogether, this makes the visitor pattern particularly expressive while preserving performance.
By design, users do not need to reason about the complexities of handling different types and may instead simply fill out the visitor implementation like a form.
This allows for the implementation of visitor logic without needing to manage reference lifetimes or specific types except when desired.

Additional traits which enable visitors to control navigation over derivation trees, such as the \texttt{visit\_each} method of \texttt{VisitableChildren} used in \cref{fig:write-visitor-impl}, have generated implementations on the opaque nodes.
Users may then constrain and invoke functions on opaque nodes to direct traversal.
Because the visitor has complete control over the traversal strategy, visitors can represent any bulk inspection or manipulation operation over derivation trees.
It is with this primitive that we achieve the remainder of this paper, whether it be evaluating constraints, mutating and crossing over derivation trees, computing diversity metrics, or any other bulk operation.

\subsection{Single- and Multi-Objective ELBT} \label{sec:evolutionary-design}

In \cref{sec:intro}, we establish that the previous state of the art, \fandango{}, struggles to accomplish certain input generation goals within a timely manner.
While our contributions so far have focused on optimizing the operations that one may take on derivation trees, there are further contributions to be made by improving the algorithms used.
In this section, we expand on just that by first analyzing the evolutionary algorithm described in \fandango{}~\cite{zamudio2025fandango}, then implementing NSGA-II~\cite{nsga2}, a classical multi-objective algorithm.

\subsubsection{The \fandango{} Evolutionary Algorithm}

The authors of \fandango{} outline the operations and algorithm used to perform ELBT in Section 3 and Algorithm 1 of the original paper~\cite{zamudio2025fandango}.
We implement the same constraint evaluation (hereon, ``check''), crossover, and mutation operations in \tool{} using the visitor pattern described in \cref{sec:visitors}.
The \fandango{} evolutionary algorithm is a very typical genetic algorithm: for user-specified parameters $p$, $e$, $c$, and $i$: (1) generate some population of size $p$ and evaluate it, (2) identify the ``elite'' (i.e. best) $e$ members of the population, (3) create the next round of $c$ candidates by mutation and crossover of the last population, (4) produce the next population from the $e$ elites and $p-e$ best candidates of generation.
Then, simply iterate 2-4 until the objectives are sufficiently satisfied (in this case: the entire population satisfies all constraints) or at iteration $i$.
Evaluation of population members is determined as a sum of the degree of satisfaction for each of the constraints specified, making this algorithm multi-objective by weighted sum.
We hereon refer to this coercion of multi-objective into single-objective as simply single-objective.
For further details, we refer the reader to Algorithm 1 of \fandango{}~\cite{zamudio2025fandango}.

\subsubsection{Implementing NSGA-II}

NSGA-II~\cite{nsga2} is an evolutionary algorithm used by \tool{}, which was originally designed specifically to optimize multiple objectives simultaneously.
Our implementation of this algorithm is effectively the same as that of the \fandango{} algorithm with two exceptions: we do not forcibly propagate elites (i.e., we propagate members of the previous population iff they are elite in the new population) and the ``best'' inputs are determined by non-dominated sorting.

To explain the purpose of non-dominated sorting, consider trying to implement a total ordering over this list of scores.
Combinators like the weighted sum used in \fandango{} suffer in that they often unfavorably advantage ``easier'' objectives and they remove nuance from the selection, leading to populations which do not ``evenly'' optimize over the objectives~\cite{weighted-sum-moea}.
Similarly, lexicographic comparison biases the search to the earlier-positioned scores, which may bias, slow, or dead-end the search by making later objectives unoptimizable over produced populations~\cite{lexicographic-optimization}.
Non-dominated sorting addresses this by sorting without implicitly favoring one objective over another.
Since domination is transitive, non-dominated sorting can be implemented efficiently with dynamic programming (so-called ``fast non-dominated sorting''~\cite{nsga2}).
In this way, we may select the cascadingly Pareto optimal candidates across all objectives.

To complete our basic NSGA-II implementation, we use the same strategy as the \fandango{} algorithm but create the next population by consuming the Pareto fronts in descending domination order.
\ifarxiv
The last consumed front, for which not all members are taken into the next population due to size limits on the population, is normally selected by approximating the most diverse set of scores, or \emph{niching}~\cite{mahfoud1995niching}.
In our implementation, users may optionally provide an alternative sorting strategy to account for their own diversity metrics, e.g., $k$-path coverage~\cite{havrikov2019systematically}.
\fi
Though there are other algorithms that likely exhibit better capabilities, we select NSGA-II because it is conceptually simple, enables multi-objective search, and demonstrates that ELBT (as implemented by \tool{}) could be relatively easily extended to other evolutionary strategies as desired.

\section{Profiling: LBT for Previously Evaluated Formats} \label{sec:profiling}

There are two ways in which we advance the state of the art for LBT: first, by optimizing the core operations, and second, by expanding the capabilities of the evolutionary approach.
We begin in this section by profiling the performance of individual operations, demonstrating the raw efficiency of our design as detailed in \cref{sec:grammar-to-rust,sec:generation,sec:visitors}.
We also perform an ablation study and analysis of various design choices taken along the way.

\subsection{Experimental Setup} \label{sec:experimental-setup}

All evaluations are performed on a server with two Intel(R) Xeon(R) Gold 6248R CPUs and 512~GiB~RAM.
The experiments are executed within an Ubuntu 22.04.5 virtual machine with 32~assigned CPUs and 256~GiB~RAM, without overbooking.
Individual experimental trials are bound to a single core with \texttt{taskset} and run one at a time.
Rust sources were built with version nightly-2025-10-09 using \texttt{-Znext-solver}\footnote{The old trait solver in the corresponding stable Rust, 1.90.0, can take several hours to build, even incrementally. The next-generation solver can build the same code within a few minutes at most. There are no perceptible runtime performance differences. We discuss the cause of this in greater detail in \cref{sec:compiler-opts-inspection}.} and Python sources were executed with Python 3.10.
The \fandango{} evaluation was completed with v1.0.0, which was the current stable release when we began this work.

\subsection{\tool{} vs. \fandango{}}
\label{sec:vs-fandango}

There are four major operations in \fandango{} we may compare against directly:
\begin{description}
    \item[Generate] produces inputs as previously described using a generator that limits the depth of the generation tree.
    \item[Check] evaluates constraints, both trivially and untrivially solvable, and returns the paths to the nodes that violate these constraints.
    \item[Mutate] traverses to a node at a specified path and replaces it with a freshly generated node.
    \item[Crossover] does the same as mutate, but instead of generating the replacement node, it generates two children by swapping the replaced node with one from another parent.
\end{description}

\noindent\textbf{Important note on reported timings:} \fandango{} computes the size of a derivation tree by the number of nonterminals and terminals, but \tool{} considers all intermediary nodes as well.
For consistency, all operations are reported in terms of their execution time \textit{per \fandango{}-style node} present in the derivation tree to which they are applied.

The code necessary to run these experiments is provided under the ``baselines'' folder of our provided source code.
Our selected targets are CSV, REST, ScriptSizeC, and XML from the original \fandango{} paper~\cite{zamudio2025fandango}.
We omit bsd-tar as this target relies heavily on explicit Python generators, for which a Rust reimplementation would be an unrepresentative advantage.
For our four chosen evaluation targets, we implement checking as visitors, preserving the semantics of the original constraints.
We use the constraints and grammars as described in the original \fandango{} paper (updated for compatibility with the version of \fandango{} used) to preserve consistency in evaluation between the papers.
We document and separately provide corrected versions of some constraints and grammars in our artifact as reference in cases where we identified issues.

We benchmark our reimplementations by direct sampling and collect profiling data from \fandango{} for each evaluated operation over a five-minute execution period.
The corresponding performance measurements are in \cref{tab:bench-results}, where operations are presented with per-node timings as determined by linear regression.
Note that, while the actual code complexity of these operations is non-linear and often not the same between tools, we use linear regression to identify \emph{principal components} of the performance impact.
For the relatively small inputs produced by \fandango{} and based on measured error, linear regression is a reasonable approximation.
The \emph{generate} and \emph{check} operations vary according to the size of the derivation tree ($n$).
\emph{Mutate} varies over both the size of the derivation tree ($n$) and the size of the regenerated (i.e., mutated) subtree ($m$).
The \emph{crossover} operation varies over the size of both parent trees ($n_1$, $n_2$) and the exchanged subtrees ($m_1$, $m_2$).

\begin{table}
    \centering
    \begin{threeparttable}
    \caption{Wall time taken to complete various operations by \tool{} and \fandango{}}    \label{tab:bench-results}
    \begin{tabular}{lrrrr}
    \toprule
     & \multicolumn{4}{c}{\tool{} ({\it \bf nanoseconds})} \\
     \cmidrule(l{0.25em}r{0.25em}){2-5}
     & \textit{Generate} & \textit{Check} & \textit{Mutate} & \textit{Crossover} \\
     \midrule
    CSV & $7.95n$ & \emph{n.d.}\tnote{1} & $0.17n + \phantom{0}8.47m$ & $1.23n_1 + 19.08n_2 - 2.44m_1 - 3.77m_2$ \\
    REST & $7.81n$ & $21.61n$ & $0.18n + \phantom{0}7.14m$ & $0.12n_1 + 20.48n_2 + 0.58m_1 + 0.35m_2$ \\
    ScriptSizeC & $12.06n$ & $31.22n$ & $0.35n + 14.04m$ & $1.36n_1 + 20.96n_2 - 4.97m_1 + 6.28m_2$ \\
    XML & $7.15n$ & $15.62n$ & $0.15n + \phantom{0}6.97m$ & $0.00n_1 + 21.19n_2 - 2.21m_1 - 6.34m_2$ \\
    \midrule
     & \multicolumn{4}{c}{\fandango{} ({\it \bf microseconds})} \\
     \cmidrule(l{0.25em}r{0.25em}){2-5}
     & \textit{Generate} & \textit{Check} & \textit{Mutate} & \textit{Crossover} \\
     \midrule
    CSV & $28.56n$ & $9.46n$ & \emph{n.d.}\tnote{1} & $0.12n_1 + 0.43n_2 + 24.27m_1 + 14.02m_2$ \\
    REST & $28.10n$ & $21.86n$ & $0.03n + 29.03m$ & $0.08n_1 - 0.32n_2 + 22.90m_1 + 24.98m_2$ \\
    ScriptSizeC & $39.52n$ & $40.64n$ & $0.19n + 39.00m$ & $0.97n_1 + 1.27n_2 + 10.91m_1 + \phantom{0}9.71m_2$ \\
    XML & $28.10n$ & $11.20n$ & $-3.88n + 97.81m$ & $-0.04n_1 + 0.53n_2 + 25.75m_1 + 18.75m_2$ \\
    \bottomrule
    \end{tabular}
    \begin{tablenotes}
        \item[1] Optimized out (\tool{}) or insufficient samples observed (\fandango{}).
    \end{tablenotes}
    \end{threeparttable}
\end{table}

Above anything else, we highlight that \tool{} measurements are made in \emph{nanoseconds} whereas \fandango{}'s are in \emph{microseconds}.
For all operations but \emph{crossover}, \tool{} has an increase in throughput of \fandangogenfactor{}.
For crossover, both \tool{} and \fandango{} are given a path to a subtree in the first parent, which is to be swapped, and scan the second parent for a subtree of equal type.
Once a candidate is identified and randomly selected, the swap takes place.
Due to its internal representation, \fandango{} must \emph{deep-copy} subtrees before swapping, but \tool{} may perform a simple byte-wise swap due to the in-memory representation.
As a result, \fandango{} primarily varies over $m_1$ and $m_2$, the size of the mutated subtrees, but \tool{} primarily varies over $n_2$, the size of the scanned parent.
Indeed, the cost is often negatively proportional to these swapped subtrees because less traversal is required (i.e., the swapped subtrees' nodes are ``higher'' in the CFG).
Though these results are only representative of the sampled distributions, \tool{} massively outpaces \fandango{} in all operations.

\conclusion{
    \tool{}'s operations are \fandangogenfactor{} faster than \fandango{}'s.
}

\subsection{Ablation Study}
\label{sec:ablation-study}

Rust is famously a ``blazingly fast'' language from the start~\cite{blazingly-fast}. If the entirety of the performance increase comes from simply using Rust, then our remaining contributions are without meaning.
To demonstrate that the optimizations we apply are indeed impactful,
we perform an ablation study across our optimizations in various contexts.

\subsubsection{Dynamic vs Static Typing}
\label{sec:vs-dynamic}

We first evaluate the performance of utilizing static typing versus dynamic typing.
We must evaluate this to justify our choice of transpilation; if a dynamically-loaded grammar is just as fast as a transpiled one, then there is no reason (at least from a performance perspective) to put in this much careful engineering effort.
Additionally, this dynamically-typed implementation is a generous facsimile of the implementation present in the original \fandango{}.
To do so, we develop a shim which allows a single node type (``DynamicNode'') to represent arbitrary grammar nodes and a sampler (``DynamicSampler'') which maintains the structure of the generated derivation tree.
This shim was extensively differentially tested against the static typing implementation to confirm equivalent structure (i.e., the same derivation tree layout) and consistent generation (i.e., generating static and dynamic inputs from the same seed produces the same input).

\begin{table}[!h]
    \centering
    \begin{threeparttable}
    \caption{Wall time of dynamic operations in nanoseconds}
    \label{tab:dynamic-results}
    \begin{tabular}{lrrrr}
    \toprule
     & \textit{Generate} & \textit{Check} & \textit{Mutate} & \textit{Crossover} \\
     \midrule
    CSV & $79.87n$ & \emph{n.d.}\tnote{2} & $1.73n + 123.26m$ & $4.32n_1 + 49.70n_2 + 1.16m_1 - 9.13m_2$ \\
    REST & $82.29n$ & \emph{n.d.}\tnote{2} & $2.50n + 101.24m$ & $2.90n_1 + 49.84n_2 + 0.16m_1 - 2.57m_2$ \\
    ScriptSizeC & $94.37n$ & \emph{n.d.}\tnote{2} & $0.51n + 138.44m$ & $4.77n_1 + 50.00n_2 - 6.75m_1 + 1.53m_2$ \\
    XML & $78.33n$ & \emph{n.d.}\tnote{2} & $2.14n + 99.62m$ & $2.45n_1 + 49.25n_2 - 3.47m_1 - 13.03m_2$ \\
    \bottomrule
    \end{tabular}
    \begin{tablenotes}
        \item[2] Unimplemented.
    \end{tablenotes}
    \end{threeparttable}
\end{table}

Our benchmarking targets for this ablation study are generation, mutation, and crossover of structured inputs, excluding checking, as the implementations vastly differed in typing.
We use the same benchmarks as \cref{sec:vs-fandango}, and present the performance of the dynamic implementations of these operations in \cref{tab:dynamic-results}.
From these results, we observe that using the dynamic version of \tool{} is inferior by up to 10$\times$ for generation, 14$\times$ for mutation, and 2$\times$ for crossover.
Crossover experiences reduced speed-up because no allocation or deallocation occurs, whereas for the other two operations, dynamic nodes must perform significantly more heap management.
Nevertheless, the performance of dynamic typing is still impressive, suggesting that it has the opportunity to address certain limitations of static typing, e.g., when the grammar dynamically changes during testing \cite{bendrissou2023grammar}.

\conclusion{
    Dynamic typing reduces performance, but has other potential applications.
}

\subsubsection{Compiler Optimizations}
\label{sec:vs-other-optimizations}

In the last section, we compare our static implementation against the dynamic implementation.
This justifies the use of the static forms of \fandango{}, but does not help us quantify the specific influence of compiler-introduced optimizations.
We instead infer this from inspecting the relationship between the static and dynamic forms when compiled without these compiler optimizations versus the same with.
This allows us to measure the degree of optimization for the respective implementations.
The results of executing the same benchmarks again with the unoptimized forms of \tool{} are shown in \cref{tab:unopt-results}.

\begin{table}
    \centering
    \begin{threeparttable}
    \caption{Wall time taken by unoptimized \tool{} in static and dynamic forms}
    \label{tab:unopt-results}    \begin{tabular}{lrrrr}
    \toprule
     & \multicolumn{4}{c}{\tool{} (static, unoptimized; nanoseconds)} \\
     \cmidrule(l{0.25em}r{0.25em}){2-5}
     & \textit{Generate} & \textit{Check} & \textit{Mutate} & \textit{Crossover} \\
     \midrule
    CSV & $195.70n$ & $0.34n$ & $3.52n + 171.02m$ & $6.10n_1 + 186.39n_2 - 9.64m_1 - 20.10m_2$ \\
    REST & $245.45n$ & $730.00n$ & $4.57n + 196.69m$ & $3.19n_1 + 234.39n_2 - 16.61m_1 - 5.57m_2$ \\
    ScriptSizeC & $275.78n$ & $675.28n$ & $6.86n + 289.74m$ & $11.69n_1 + 200.59n_2 - 15.21m_1 + 7.91m_2$ \\
    XML & $197.69n$ & $308.79n$ & $3.31n + 178.92m$ & $4.55n_1 + 199.88n_2 - 15.56m_1 - 37.96m_2$ \\
    \midrule
     & \multicolumn{4}{c}{\tool{} (dynamic, unoptimized; nanoseconds)} \\
     \cmidrule(l{0.25em}r{0.25em}){2-5}
     & \textit{Generate} & \textit{Check} & \textit{Mutate} & \textit{Crossover} \\
     \midrule
    CSV & $704.27n$ & \emph{n.d.}\tnote{2} & $4.50n + 783.36m$ & $10.72n_1 + 517.57n_2 - 12.10m_1 - 22.84m_2$ \\
    REST & $709.50n$ & \emph{n.d.}\tnote{2} & $4.90n + 780.17m$ & $3.12n_1 + 563.43n_2 + 0.21m_1 - 9.71m_2$ \\
    ScriptSizeC & $753.57n$ & \emph{n.d.}\tnote{2} & $6.36n + 840.72m$ & $8.82n_1 + 450.56n_2 - 21.89m_1 + 6.10m_2$ \\
    XML & $672.43n$ & \emph{n.d.}\tnote{2} & $5.10n + 725.91m$ & $-0.23n_1 + 523.14n_2 - 13.31m_1 - 6.21m_2$ \\
    \bottomrule
    \end{tabular}
    \begin{tablenotes}
        \item[2] Unimplemented.
    \end{tablenotes}
    \end{threeparttable}
\end{table}

We observe that the unoptimized form of the static implementation suffers a performance penalty of factors 30, 27, and 11 for principal parameters of generation, mutation, and crossover, respectively.
The dynamic implementation, on the other hand, suffers only 9, 8, and 11, respectively.
We infer from this that the effect of optimizations on the static implementation is roughly three times larger for operations where the primary cost factor is grammar-dependent (i.e., excluding crossover, as there is no structural advantage in its implementation).
In other words, the Rust compiler has three times the capacity to optimize the static implementations versus the dynamic implementations when structure is a determining factor of optimizability, demonstrating that the compiler is significantly better at optimizing these grammar operations when formatted statically.
How the compiler actually optimizes this is explored further in \cref{sec:compiler-opts-inspection}.

\conclusion{
    Rust compiler optimizations are more effective on statically-typed grammars than dynamically-typed grammars.
}

\subsection{Inspection of Compiler Optimizations} \label{sec:compiler-opts-inspection}

Some parts of our design cannot be tested in an ablation study, e.g. \cref{sec:generation,sec:visitors}, because these are core design elements of our implementation.
It is therefore difficult to evaluate the specific impact that these design choices made compared to that of our structural choices.
By inspecting the produced executables, we can observe how our design choices allow the compiler to make impressive inferences.

\begin{figure}
    \centering
    \small
    \begin{cminted}{rust}
fn visit<'program, N>(mut self, node: &'program N, _idx: usize)
-> VisitResult<Self, T>
where /* ... boilerplate constraints ... */
{
    // check the definition of this node in the grammar
    match node.definition() {
        // if this is a nonterminal or a terminal: count it!
        FandangoNode::Nonterminal(_) | FandangoNode::String(_) => {
            self.count += 1;
        }
        _ => {} // otherwise, do nothing
    }
    node.opaque().visit_each(self) // recurse
}
    \end{cminted}
    \caption{A snippet from our \texttt{visit} method implementation for \texttt{FandangoNodeCounter}.}
    \label{fig:counter-snippet}
\end{figure}

Consider the snippet in \cref{fig:counter-snippet} which implements the node counting strategy described at the start of \cref{sec:vs-fandango}.
This visitor implementation scans the whole derivation tree by pre-order traversal and counts the number of nonterminals or terminals, and thus enables our per-node comparison against \fandango{}.
One might expect that, because we must check each node for this condition, the cost of this visitor is exactly the size of the derivation tree.
In reality, the compiler does something far more clever.
The \texttt{definition} method on \texttt{node} returns a constant value: the corresponding definition of the derivation tree in the grammar.
Because the compiler observes that this is constant, it can infer the result of the match, therefore concretizing count here to either just an increment or a no-op followed by recursion.
This is only possible because the \texttt{visit} function itself is generic over \texttt{N}, the node type itself.
If \texttt{visit} was instead generic over e.g. the opaque type, the contained node type would first need to be determined by another match.

Next, we recurse across each child node.
In the source code, this first upcasts \texttt{N} to its opaque type \texttt{T}, then invokes the \texttt{visit\_each} method of \texttt{T}.
This method determines what concrete type it contains (trivially, \texttt{N}), then iterates over the children of \texttt{node} in order using \texttt{N}'s \texttt{visit\_each}\footnote{This small dance is necessary because we cannot constrain \texttt{N} at the function generic. If we defined visitors as generic over \texttt{N} instead of \texttt{T}, we would instead have issues with e.g. ensuring return value consistency or cyclic trait requirements when implementing \texttt{N}'s \texttt{visit\_with}. We balance this by giving the compiler enough information to optimize this detail away.}.
When compiled, the upcast/downcast pair is trivially resolved and thus elided, such that the iteration takes place directly.
The subsequent calls to the same visit function, but with a different concretization of~\texttt{N}, may then be inlined---aggressively.
\cref{fig:csv-optimization} demonstrates this degree of optimization on the CSV grammar, for which the corresponding grammar snippet is provided in \cref{fig:csv-snippet-opt}.

\begin{figure}
    \centering
    \begin{minipage}{0.45\textwidth}
    \begin{densegrammar}
<csv\_file> ::= <csv\_header> <csv\_records>

<csv\_records> ::= <csv\_record> <csv\_records> | `'
    \end{densegrammar}
    \end{minipage}
    \caption{Grammar snippet of CSV relevant for \cref{fig:csv-optimization}.}
    \label{fig:csv-snippet-opt}
\end{figure}

\begin{figure}
    \centering
    \small
    \begin{cminted}{c}
Result<> __rustcall
fandango_targets::csv::defs::__Csv_defs::visit_each<>
          (nonterminal_csv_file_0 *self,FandangoNodeCounter visitor)
{
  /* ... variable decls ... */
  uVar1 = /* ... large inlined traversal ... */;
  if (uVar1 == 96) { // check discriminant for a trivial variant
    RVar3 = (Result<>)(visitor.count + 10);
  } else { /* ... other cases ... */ }
  nVar11 = (self->child_1).child_0;
  // descend csv_records until the right expansion
  while (nVar11 != (nonterminal_csv_records_0)0x0) {
    // variant of the contained csv_record
    uVar2 = *(uint8_t *)((long)nVar10 + 0x10); 
    if (uVar2 == 96) { // trivial variant check again
      RVar3 = (Result<>)((long)RVar3 + 11); // one extra intermediate!
      // continue recursion by descending the csv_records pointer
      nVar11 = *(nonterminal_csv_records_0 *)nVar11;
    } else { /* ... other cases ... */ }
  }
  return (Result<>)((long)RVar3 + 3);
}
    \end{cminted}
    \caption{Ghidra~\cite{ghidra} decompiler snippet for \texttt{visit\_each} for the CSV grammar's \texttt{csv\_file} expansion.}
    \label{fig:csv-optimization}
\end{figure}

The Rust compiler makes several optimizations when inlining these functions, like recursive to iterative transformation, merging of increment operations, struct layout optimizations, inferring minimum subtree sizes, etc.
Our design decisions in the visitor pattern are symbiotic with our choices in static typing, and we observe similar optimizations for samplers and generators, which are monomorphized similarly.
In short: every design decision presented in \cref{sec:design} is carefully chosen to provide the maximum amount of information to the compiler about the structure of the data, which the compiler then uses to produce maximally optimized operations from user-defined code like that of \cref{fig:csv-optimization}.
Indeed, we find optimal implementations of \texttt{visit\_each} like this for every \texttt{N} such that when we invoke the visitor on a non-trivial \texttt{T} (e.g., to measure the size of a mutated subtree), we will always execute a optimum implementation.
Dynamic nodes, conversely, cannot be so optimized because the structural information is simply not present.

\conclusion{
    Our static transformation uniquely enables the Rust compiler to optimize downstream user code.
}

This optimization does not ``come for free'', so to speak.
The compiler must optimize over $O(k^n)$ paths during monomorphization, with $k$ as the branching factor of the grammar's graph.
Without the use of \texttt{-Znext-solver} (otherwise known as the ``next-generation trait solver''), these paths are traversed and cached repeatedly, leading to extreme compilation times and memory consumption; the REST grammar, for example, can take several hours and over 100 GiB of RAM to compile.
Conversely, with \texttt{-Znext-solver}, trait implementations are computed and cached more optimally, reducing compilation time and memory usage to those of typical projects (a few minutes and <2 GiB RAM, respectively)~\cite{next-solver}.
While not evaluated formally in this paper, this alternate solver allowed us to scale to very large grammars with relatively high branching factors.
Without this, our work would likely not be practical.

\section{Case Study: LBT for Compiler Testing} \label{sec:case-study}

The performance improvements (particularly for constraint checking) afforded by our transformation can allow LBT to be applied in domains where traditional approaches struggle, and what better target for highly constrained input generation than compiler testing?
Indeed, this was the context in which grammar-based test generation was first explored~\cite{burkhardt1967}.
When generating programs that are supposed to reach deep into a compiler, it is critical to get a number of things right from the beginning.
For instance, one does not want to generate programs where undeclared or uninitialized variables are used, or where non-existent fields of objects or structures are accessed.

In this case study, we measure along two baselines: against the state of the art, \fandango{}, and against \tool{} using \fandango{}'s evolutionary algorithm.
One would rightly point out that comparing a Rust implementation to a Python implementation is comparing languages of inherently different speeds.
To demonstrate that our contributions go beyond the choice of language, the second part of this case study inspects the performance of our multi-objective approach versus the single-objective approach present in \fandango{}.
That is: \tool{}, including typing optimizations, in a multi-objective and a single-objective configuration.
The single-objective configuration of \tool{}, which uses the algorithm described in the original \fandango{} paper~\cite{zamudio2025fandango}, serves as a very conservative upper bound for the performance of a Rust-based reimplementation of the original \fandango{} strategy, given the dynamic- versus static-typing results of \cref{sec:vs-dynamic}.

\subsection{Experimental Setup}
\label{sec:compiler-test:experimental-setup}

We conduct a small case study demonstrating the potential of \tool{} in the context of compiler testing.
For this case study, we consider a subset of the C grammar without conditional statements, arrays, and pointers.
We identify the following (non-exhaustive) list of constraints that should be valid for generated programs:
\begin{multicols}{2}
\begin{itemize}
    \item variables declared before use;
    \item no empty struct definitions;
    \item no access to non-existent struct fields;
    \item no void types outside of function returns;
    \item functions with non-void returns should return a value;
    \item no duplicate struct field names;
    \item no re-declarations in the same scope;
    \item no identifiers should be reserved keywords;
    \item program should be well-typed.
\end{itemize}
\end{multicols}

We implemented these constraints with visitors in \tool{} and similarly in \fandango{}, then performed an experiment wherein we had \tool{} and \fandango{} generate programs that satisfied these constraints.
This experiment used the same hardware configuration as in \cref{sec:experimental-setup}.
Since these are not necessarily exhaustive constraints, we passed each produced input to the GNU Compiler Collection (gcc)~\cite{gcc} C compiler (version 11.4.0) to check for validity.
In this process, we tracked how many valid programs each tool generated in one minute, as well as the diversity of the generated programs; to measure diversity, we computed the overall $k$-path coverage of the valid inputs generated in the time limit.
Both tools were configured to generate an initial population of 100 and then evolve that population until meeting their termination conditions repeatedly until over one minute total of generation and compilation elapsed\footnote{Here, we execute until the one minute is surpassed and the generation process terminates cleanly.}.
We perform this experiment five times for each configuration and report the mean over these trials.

\subsection{Results}

The results of our experiment are summarized in \cref{tab:compiler-test-summary}.

\begin{table}
    \centering
    \caption{Results of \tool{} and \fandango{} generating C programs until one minute is surpassed.
    }
    \label{tab:compiler-test-summary}
    \begin{threeparttable}
    \begin{tabular}{ccrrrrr}
        \toprule
         & \multicolumn{6}{c}{\tool{}} \\
          \cmidrule(l{0.25em}r{0.25em}){2-7}
        Objectives & Multi? & \# Valid & $5$-path & Gen. Time & {\tt gcc} Time & Total \\
        \midrule
        $\varnothing$ & N/A & \unconstrainedCompileAvgRs/\unconstrainedTotalAvgRs & \unconstrainedKPathCoverageAvgRs & \unconstrainedGenTimeAvgRs & \unconstrainedCompileTimeAvgRs & \unconstrainedTotalTimeAvgRs \\
        \multirow{2}*{Validity} & \xmark & \singleValidCompileAvgRs/\singleValidTotalAvgRs & \singleValidKPathCoverageAvgRs & \singleValidGenTimeAvgRs & \singleValidCompileTimeAvgRs & \singleValidTotalTimeAvgRs \\
         & \cmark & \multiValidCompileAvgRs/\multiValidTotalAvgRs & \multiValidKPathCoverageAvgRs & \multiValidGenTimeAvgRs & \multiValidCompileTimeAvgRs & \multiValidTotalTimeAvgRs \\
        \multirow{2}*{Validity $\cup$ Generation} & \xmark & \singleValidAndSizedCompileAvgRs/\singleValidAndSizedTotalAvgRs & \singleValidAndSizedKPathCoverageAvgRs & \singleValidAndSizedGenTimeAvgRs & \singleValidAndSizedCompileTimeAvgRs & \singleValidAndSizedTotalTimeAvgRs \\
         & \cmark & \multiValidAndSizedCompileAvgRs/\multiValidAndSizedTotalAvgRs & \multiValidAndSizedKPathCoverageAvgRs & \multiValidAndSizedGenTimeAvgRs & \multiValidAndSizedCompileTimeAvgRs & \multiValidAndSizedTotalTimeAvgRs \\
        \midrule
         & \multicolumn{6}{c}{\fandango{}} \\
          \cmidrule(l{0.25em}r{0.25em}){2-7}
        Objectives & Multi? & \# Valid & $5$-path & Gen. Time & {\tt gcc} Time & Total \\
        \midrule
        $\varnothing$ & N/A & \unconstrainedCompileAvgPy/\unconstrainedTotalAvgPy & \unconstrainedKPathCoverageAvgPy & \unconstrainedGenTimeAvgPy & \unconstrainedCompileTimeAvgPy & \unconstrainedTotalTimeAvgPy \\
        Validity & \xmark & \validConstrainedCompileAvgPy/\validConstrainedTotalAvgPy & \validConstrainedKPathCoverageAvgPy & \validConstrainedGenTimeAvgPy & \validConstrainedCompileTimeAvgPy & \validConstrainedTotalTimeAvgPy \\
        Validity $\cup$ Generation & \xmark & \textit{n.d.}\tnote{3} & \textit{n.d.}\tnote{3} & \textit{n.d.}\tnote{3} & \textit{n.d.}\tnote{3} & \textit{n.d.}\tnote{4} \\
        \bottomrule
    \end{tabular}
    \begin{tablenotes}
        \item[3] Unable to produce any inputs.
        \item[4] Process self-aborted before producing any inputs.
    \end{tablenotes}
    \end{threeparttable}
\end{table}

\subsubsection{Totally Unconstrained}

When generation is unconstrained, the majority of the one-minute period is spent on compilation (i.e., actually executing the generated inputs to the system under test).
Because evolution is not needed, the multi-objective and single-objective implementations of \tool{} are actually identical for this case.
One can see that \tool{} is much faster at generating the actual inputs than \fandango{}, with \tool{} being around 350x faster than \fandango{} in input production.
\fandango{} generates more valid programs because its generator favors trivial inputs, whereas \tool{} does not.
As a result, we observe this difference in acceptance rate: longer, random inputs are less likely to be valid.

\subsubsection{Adding Validity Constraints}
Since compilation is the bottleneck, let us add validity constraints to both specifications.
We configured both tools with the constraints mentioned in \cref{sec:compiler-test:experimental-setup}.

We similarly see that \fandango{} begins to slow down here dramatically, spending nearly the entire allotted time on input generation.
Validity requirements force both tools to search for inputs that satisfy requirements.
As a result, program generation takes longer, but the subset of valid inputs (though greater by percent from the unconstrained experiment) decreases in size.
Without specific guidance to produce diverse or larger inputs, the selection of paths in a single input is due largely to luck\footnote{This is an instance of the coupon collector's problem~\cite{coupon-collector}.}.
For these reasons, we see a decrease in \fandango{}'s diversity in this experiment directly related to the decrease in the number of inputs executed.

Overall, we see that both forms of \tool outperform \fandango; it generates more valid programs in far less time.
Because vacuous satisfaction of constraints is simpler, all strategies evolve their populations into \textit{very small programs}, with at most a couple of statements; \cref{lst:validity_constrained_examples} shows a few of the more interesting generated programs from this phase of the experiment, but all are roughly of that size or smaller.

\begin{figure}[!h]
 \begin{minipage}{0.49\textwidth}
 \centering
 \small
  \begin{cminted}{c}
int Udm = {
  true
};
  \end{cminted}
 \end{minipage}
 \hfill
 \begin{minipage}{0.49\textwidth}
 \centering
 \small
  \begin{cminted}{c}
double wz( ) { 
  return 0x5; 
}
  \end{cminted}
 \end{minipage}
 \caption{Examples of programs generated by \fandango{} and \tool{} with validity constraints.
 }
  \label{lst:validity_constrained_examples}
\end{figure}

\subsubsection{Adding Generation Objectives}
To address the low complexity of inputs, we add a few constraints to help create larger inputs.
Here, we require that each program have at least \sizeConstraintMinStatements statements, with at most \sizeConstraintStructLimit struct definition and \sizeConstraintFunLimit function definition (as these are relatively unconstrained as compared with, e.g., expressions, which need to be well-typed).

\begin{figure}[!h]
 \begin{minipage}{0.49\textwidth}
  \centering
  \small
  \begin{cminted}{c}
void wE(double e, char R) { 
  int G;
  bool SdF6 = 15;
  !true;
  !true;
  return;
}
float j2;
  \end{cminted}
 \end{minipage}
 \hfill
 \begin{minipage}{0.49\textwidth}
  \centering
  \small
  \begin{cminted}{c}
float G = 9.7f;
double N8( ) { 
  float A;
  return '2';
  char m;
  return 'R';
  int L;
}
int L;
  \end{cminted}
 \end{minipage}
 \caption{Examples of programs generated by \tool with validity and size constraints.}
  \label{lst:validity_and_size_constrained_examples}
\end{figure}

Multi-objective \tool generated \multiValidAndSizedTotalAvgRs programs on average, and a majority of these passed a call to \texttt{gcc}.
The programs generated with this configuration are small and still favor vacuous satisfaction of constraints, though do so with larger programs; consider \cref{lst:validity_and_size_constrained_examples} for some examples.
In contrast to this, \fandango was incapable of generating a single valid input in the allotted time.
Indeed, \fandango{}'s EA detects that it is failing to progress in these constraints and increases the size of the population to compensate.
Letting it run for a period significantly longer than the time allotted eventually causes the process to abort, finding no inputs.
Single-objective \tool{}, which acts here as an upper bound for the performance of a Rust implementation of \fandango{}, also produced no valid inputs.
This demonstrates the importance of the multi-objective algorithm predicted by EA literature~\cite{nsga2}: the single-objective approach gets caught in local optima when producing inputs, as increasing the satisfaction of size constraints without reducing the satisfaction of other constraints is difficult.
Probabilistically, the single-objective approach is unlikely to produce non-trivial inputs at a reasonable rate, regardless of the choice of language.
Similarly, based on the results of \cref{sec:vs-dynamic}, we infer that we would produce inputs at one-tenth of the rate with a dynamic, multi-objective version of \tool{}.
In other words: both the optimization and the choice of algorithm is important for the high throughput of complex input generation.

We can also specify a more general generation objective, such as a ``node goal'' stating that at least, e.g., 500 nodes should be present in generated solutions.
\tool can still generate inputs like this, although this generation objective takes longer to solve; consider \cref{lst:validity_and_size_with_node_goal_example} for some examples of programs generated in this way instead.

\begin{figure}
    \centering
    \small
    \begin{minipage}{0.42\textwidth}
    \begin{cminted}{c}
    char b03(bool o) { 
      0xC;
      char s6d3qU80Skqs10H = 27 != 626.5;
      float vPh4 = 27 != -749.138;
      return -749.138; 
    }
    float f2MY( ) { 
      struct r45h6 {
        char lA;
        float xYXaIt;
        float oz14V;
      };
      char s1v042s = 27 != 627.5;
      return -749.138; 
    }
    float c3;    
    \end{cminted}
    \end{minipage}
    \hfill
    \begin{minipage}{0.52\textwidth}
    \begin{cminted}{c}
    void Wx(int O9mC) { 
        int OD = 8.496f * 69205.51252;
        return;
        474870.6552304881f;
        double rEnNO5U; 
    }
    bool ui6;
    void Tj(int Y16) { 
        int OD = 474870.33f || 69205.2;
        return;
        int IJ3x99 = '8';
        double T;
    }
    
    int hp = 78923789237 
    void  US( ) { return; }
    \end{cminted}
    \end{minipage}
    \caption{Examples of larger programs generated by \tool.}
    \label{lst:validity_and_size_with_node_goal_example}
\end{figure}

\subsection{Summary of Case Study}

With just a grammar, a few constraints, and generation goals, our design of \tool{} allows one to generate large, complex test inputs in a matter of seconds.
On the one hand, performance improvements based on our transformation allow for performant generation, constraint evaluation, and input modification, and on the other, our multi-objective evolutionary strategy allows for many constraints and generation objectives to be solved in tandem, and efficiently.
Most importantly, we demonstrate by ablation that code optimization alone is not enough; the presence of multiple high complexity constraints demands the use of evolutionary algorithms more suited to solving multiple orthogonal objectives simultaneously, as predicted by EA literature~\cite{nsga2}.

\subsection{Finding Bugs in the Claude C Compiler}

Anthropic recently reported that Claude Opus 4.6 created a compiler for C, called \texttt{CCC}, which received significant attention~\cite{ccc-compiler}.
We crafted a small \textit{differential testing} scenario, pitting \texttt{CCC} against \texttt{gcc}, to see if our approach described here could detect any differences. 
We ran \tool for one minute, as above, but in order to detect if one compiler compiled programs that \textit{should not compile}, we also invoked both compilers on \textit{almost correct programs}, i.e., generated programs that had a few constraint violations. 
This is another advantage of our approach: fitness is a measure of correctness, so we know how close a program is to being semantically correct (at least, with respect to our constraints).
In this short experiment, we found two notable discrepancies, discussed below.
Note that other compiler testing tools, like Csmith~\cite{yang2011csmith} and \texttt{yarpgen}~\cite{yarpgen}, also found bugs in \texttt{CCC}~\cite{csmith-ccc-problems}, though these bugs are mutually exclusive from our own.

\begin{figure}
    \centering
    \small
    \begin{minipage}{0.42\textwidth}
    \begin{cminted}{c}
    int e = 591;
    int e = 035502;
    \end{cminted}
    \subcaption{\texttt{CCC} allows same-scope redeclarations.}
    \end{minipage}
    \hfill
    \begin{minipage}{0.52\textwidth}
    \begin{cminted}{c}
    double Z = 'a';
    int x7 = Z;
    \end{cminted}
    \subcaption{\texttt{CCC} allows non-constant global variable initialization.}
    \end{minipage}
    \caption{Exemplary snippets of \texttt{CCC} mis-compilations generated by \tool;
    these are reduced from larger generated programs for illustrative purposes.
    \texttt{gcc} rejects both of these programs.}
    \label{lst:ccc-miscomp}
\end{figure}

\subsubsection{Difference 1: \texttt{CCC} silently allows redefinitions}

Same-scope redefinitions of variables are disallowed in C; \texttt{gcc} enforces this, but \texttt{CCC} does not, silently ignoring the redefinition.
See \cref{lst:ccc-miscomp} (a) for an exemplary snipped drawn from a generated program showcasing this difference.

\subsubsection{Difference 2: \texttt{CCC} allows non-constant global variable initialization}

Global variables in C should be initialized with constant expressions only;
we found that this is enforced in \texttt{gcc}, but not in \texttt{CCC}. 
See \cref{lst:ccc-miscomp} (b) for an exemplar generated snippet.

\section{Threats to Validity} \label{sec:threats-to-validity}

Our experiments are subject to a number of threats to validity.

\begin{description}
\item[Internal validity.] As with all stochastic search-based techniques, \tool{}'s performance and coverage are subject to variation across executions. 
    While we performed multiple runs to mitigate noise, the random nature of genetic algorithms may cause differences in result quality.
    Additionally, the performance analysis relies on observed metrics such as execution time and input validity, which are influenced by implementation details and the system configuration used for evaluation.
    Our measurements are thus representative of this specific setup and should not be assumed to generalize without replication.
\item [External validity.] Our comparison is based on four input formats that mirror those used in the original \fandango{} study.
    While this enables direct comparison, it also limits the breadth of formats considered.
    As noted in the original \fandango{} work~\cite{zamudio2025fandango}, these formats may not capture the full diversity of real-world input grammars, particularly in domains such as multimedia, compressed formats, or domain-specific languages.
    Consequently, our findings may not extend to grammars with substantially different structures, constraints, or complexity levels.
    In a similar vein, our case study (\cref{sec:case-study}) may or may not generalize to other programming languages or more complex C features.
\item [Construct validity.] This work builds directly on the conceptual model of \fandango{}, inheriting both the strengths and limitations of evolutionary language-based testing.
\end{description}

\section{Limitations} \label{sec:limitations}

Like \fandango{}, we have a few theoretical limitations.
\begin{description}
    \item[Unknown Unsatisfiability.] Evolutionary algorithms have no means by which to determine the unsatisfiability of constraint satisfaction problems (CFPs), and thus ELBT cannot ``reject'' such cases.
    While \tool{}'s implementation of NSGA-II allows us to satisfy maximally, it cannot allow us to determine unsatisfiability.
    This is especially problematic in the presence of quantifier logic that might make the constraints solvable only under conditions not yet represented in the population.
    Indeed, the theoretical limits of EA applied to quantified CSPs, like those used in our grammars, are not yet well-explored.
\item[Need Guidance, or Crossover] Similarly, EAs work best when satisfaction of a constraint is possible by crossover (e.g., def-use patterns in \cref{sec:case-study}) or when there is a gradient (e.g., integer equality has a ``distance to satisfaction'').
When neither of these is present, the EAs must fall back on luck---for which the probability of solving might be so low that even our optimizations cannot help us.
\item[Inherited Limitations] Otherwise, the classic limitations of EAs apply: stochastic search is not guaranteed to find globally optimal solutions nor any solution at all for the reasons above and more.
\end{description}

\section{Related Work} \label{sec:related-work}

\subsection{Random Testing}

At the core of our work is \textit{random testing}, or generating random inputs and passing them to a system under test.
This simple concept is actually integral to many fields of research, the most related being \textit{fuzzing}, \textit{search-based testing}, and \textit{property-based testing}.
We begin by sketching the origins of these fields, then allude to the research directions they subsequently followed, and end with a highlight of their many technical and academic similarities.

\subsubsection{(Grammar-Based) Fuzzing}

Fuzzing, as introduced by Miller et al.~\cite{miller1990empirical}, is a process where random byte streams are generated and used to test the reliability of system utilities\footnote{
Arguably, Breuer in 1971~\cite{breuer1971algorithm} and Purdom in 1972~\cite{purdom1972} precede this, but Miller et al. coined the term ``fuzzing''.
}.
The fuzzing community most often treats the target, or system under test, as a \textit{black-box} binary that they know nothing about, and the primary impetus of research has been to come up with strategies that assume as little as possible about said target.
Most related to our work is grammar-based fuzzing, and this basic principle of using grammars to produce test inputs dates back to
Burkhardt~\cite{burkhardt1967} in 1967 (making it the oldest systematic test input generation technique), to be later rediscovered by Hanford~\cite{hanford1970} and Purdom~\cite{purdom1972}, who used grammars of Algol, FORTRAN IV, and a major subset of PL/I to produce test inputs in these languages, though these works focused on testing the parsers.
These grammars act as system-level specifications, but the system is often still opaque: e.g., without knowing how an HTML parser is implemented, an HTML grammar is still helpful for testing.

\subsubsection{Search-Based Testing}

Search-based software testing (SBST) is a field where metaheuristic optimization algorithms are used to automatically generate test cases for systems.
Rather than writing specific, tailored input generators, SBST users specify a fitness function that says how good a given input is, and then the SBST algorithm will search for the fittest inputs.
SBST dates back to 1976 in the work of Miller and Spooner~\cite{miller1976automatic}, who used numerical maximization to generate floating-point data, essentially prioritizing inputs that were ``closer'' to executing a desired path through a program.
Then, in the early 1990s, Korel~\cite{korel1990automated, korel1992dynamic} revisited the problem, e.g., exploring how incorporating dynamic analysis affects performance, and the field really came into its own with Xanthakis et al. in 1992~\cite{xanthakis1992application} when they proposed using genetic algorithms for maximizing fitness, setting the stage for the vast majority of subsequent research.
When SBST approaches leverage execution information to compute fitness, they are \textit{grey-box} approaches, typically relying on instrumentation for feedback but otherwise being ignorant of the code.

\subsubsection{Property-Based Testing}

Property-based testing (PBT), pioneered by QuickCheck~\cite{claessen2000quickcheck}, tests a given property of a program by generating inputs and asserting the property on them.
The scope of PBT has expanded much beyond testing properties of pure functions written in a functional language~\cite{padhye2019jqf, yatoh2014arbitcheck, reddy2020quickly, goldstein2024property} (as was the case for QuickCheck), but still most opt for a \textit{white-box} approach where the tester has full knowledge of the system under test; the traditional use-case for PBT is for a user to test their own system, rather than test other, unknown systems.
Indeed, inputs are usually generated according to in-language data structures.

\subsection{Synthesis}

Fundamentally, fuzzing, SBST, and PBT are all flavours of random testing.
These fields are much more similar than they are distinct, exemplified by how they suffer and solve the same problems.
For instance, the diversity of generated inputs is a concern for all fields, and is being actively explored in fuzzing~\cite{srivastava2021gramatron, nguyen2022bedivfuzz}, PBT~\cite{claessen2015generating, boyapati2002korat}, and SBST~\cite{albunian2017diversity, albunian2020measuring}.
Also, generating inputs with many constraints (i.e., satisfying sparse preconditions) is always a challenge, also being actively explored in fuzzing~\cite{godefroid2008grammar, eberlein2020evolutionary, aschermann2019nautilus, hodovan2018grammarinator}, PBT~\cite{lampropoulos2019coverage, lampropoulos2017generating}, and SBST~\cite{zhu2019improving, malburg2011combining}.

We have also reached a point where each of these research areas is quite mature and, we argue, are now borrowing from each other for inspiration.
Guiding input generation with coverage in fuzzing (e.g.,~\cite{aschermann2019nautilus}) and PBT (e.g.,~\cite{lampropoulos2019coverage, reddy2020quickly, padhye2019jqf}) is similar to the idea of ranking and maximizing input fitness from SBST; the key observation here is that black- and white-box approaches do benefit greatly from grey-box instrumentation and feedback.
In enhancing grammars with explicit pre-conditions, language-based testing~\cite{zamudio2025fandango, steinhofel2022input} starts to bridge the gap between PBT and grammar-based testing, and our work in which we compile grammar-based specification into a Rust data structure builds on this further.
Here, language-based testing is trying to build rich input specifications and squeeze as much as possible form them, which is in line with PBT work on making best use of existing rich and complex data structures.
Entire classes of fuzzing approaches, e.g., those falling under the umbrella of \textit{metamorphic testing}~\cite{chen2018metamorphic, segura2016survey} which specifies relations relating equivalent inputs, or in incorporating system properties as oracles in fuzzing~\cite{xie2022rozz, sun2023property}, also sit somewhere between fuzzing and PBT.

\subsubsection{Where does \tool stand?}

\begin{figure}
    \centering
    \small
    \begin{cminted}{rust}
fn do_another(x: &str) -> bool {
  let mut v = 0;
  if x[0..1] == "a" {
    v += 1;
  }
  if x[1..2] == "b" {
    v += 1;
  }
  /* 8 additional cases */
  v == 10
}
    \end{cminted}
    \caption{A function difficult to search via mutation for a truthy result by both edge and path coverage~\cite{crump39c3}.}
    \label{fig:edge-guidance-counterexample}
\end{figure}

To help establish our place in the literature, first, we argue that our work is thematically similar to coverage-guided, property-based testing~\cite{lampropoulos2019coverage} (CGPT).
Geared towards scenarios with sparse preconditions, CGPT augments PBT with ideas borrowed from mutation-based testing, coverage-guided fuzzing, and SBST; instead of discarding inputs that do not satisfy preconditions, CGPT will mutate inputs that achieve higher precondition and system path coverage.
CGPT generates mutators, and many of the mutations are similar to those in ELBT:
for instance, mutation in \fandango{} and \tool{} picks a random sub-tree of a derivation tree and re-generates it; which corresponds to applying the recursive mutation operator in CGPT, eventually replacing some element with a value of the appropriate type.
On the other hand, CGPT may not be able to infer sufficient guidance from path coverage alone.
In the fuzzing community, path coverage has seen little adoption due to the path explosion problem~\cite{pathafl,prudent-practices,path-fuzzer}.
Furthermore, there are degenerative cases where path coverage will probabilistically explore large spaces without progress towards satisfaction; see \cref{fig:edge-guidance-counterexample}, a problem known in fuzzing folklore~\cite{folklore} to be difficult to explore by edge- and path-coverage, but easy to satisfy by maximizing the variable \texttt{v} via a custom fitness function.
Works like IJON~\cite{ijon} make it possible for fuzzers to guide exploration of a program with simple variable monitoring via user annotation, but not all state variables are so trivially exposed---and sometimes, the source code of a program is not available at all.
What we show with this work is that (a) multi-objective guidance is important for complex problem domains, and (b) emitting data structures as we describe yields substantial improvements in generation throughput, as well as precondition computation.
Our work indirectly contributes to PBT and directly to fuzzing by decoupling generation from program implementation, allowing users to specify generation objectives (including constraints), fitness functions, and custom mutation operators which account for semantics beyond input and program structure.

\subsection{Compiler Testing}

In our work, compiler testing is presented as a case study to showcase where the throughput gains of \tool can be appreciated, namely in an environment with many constraints on input validity, a central challenge in testing domains with sparse preconditions.
This is a well-studied area~\cite{chen2020survey}, and the seminal work is, of course, Csmith~\cite{yang2011csmith}:
a heroic multi-year development effort, Csmith yields syntactically and semantically valid C code free of undefined behavior and with a singular interpretation, allowing programs to be compiled with various compilers (or compiler settings) and compare whether the generated programs produce the same results.
Csmith generates programs top-down by generating one statement at a time, from a list of allowable statements at the current generation point, constructing valid statements according to a context maintained through ongoing program analysis by the generator.
To be clear, Csmith is a much more complete C compiler testing tool than what we presented in \cref{sec:case-study}.

In contrast to Csmith's approach, ours is declarative: given a grammar and a set of constraints over semantic validity, we show that there is promise in relying on an evolutionary algorithm to guide generation towards valid programs.
Grammars have been used as the basis for compiler testing, as early as the work of Purdom~\cite{purdom1972} (aimed at testing parsers), through LangFuzz~\cite{holler2012fuzzing} (using grammars to remix existing code fragments), and more recently Nautilus~\cite{aschermann2019nautilus} (grammars + coverage guidance).
That said, most recent work is underpinned by hand-written generators targeting specific features, or otherwise engineered to generate correct programs~\cite{chaliasos2022finding, zang2022compiler, sharma2023rustsmith}; this is sensible, as semantically complex programs are difficult to generate purely randomly.
That said, these manually tuned generators are inherently biased, and there has been some effort on weakening the validity of generated programs in a controlled manner~\cite{even2022csmithedge} to address this bias.
There is, of course, no shortage of effort in using machine learning and large language models for compiler testing~\cite{ye2023generative, yang2024whitefox, gu2023llm}, but such approaches are most effective at testing established languages with lots of code to learn from; in contrast, an approach using only a grammar and constraints can be quickly instantiated to test new, emerging languages.

\section{Conclusion and Future Work} \label{sec:conclusion}

We present a novel approach to evolutionary language-based testing that is \fandangogenfactor{} faster than the state-of-the-art in its operations and more robust in its choice of evolutionary algorithm, allowing us to apply specification-based test generation in scenarios where it was previously infeasible.
We achieve this performance boost by transforming grammars and implementing constraints into highly optimizable Rust code, notably using data types that efficiently support the better evolutionary algorithms chosen for solving constraints.
As a consequence, we elevate specification-based test generation to much more complex domains:
In our case study, our \tool prototype produced diverse and complex C code with high speed and high precision.

Besides general improvements, our future work will focus on the following topics:

\begin{description}
    \item[Parameterized compiler testing.] Our C case study demonstrates that a grammar and a small set of constraints suffice to generate an arbitrarily large and diverse set of code inputs, of which the vast majority are valid.
    Given that
\begin{enumerate*}[label=(\arabic*)]
    \item pretty much every programming language comes with a formal grammar, and
    \item the constraints we formulated for C can be easily adapted to other languages,
\end{enumerate*}
    \tool brings new promises to automated compiler testing, especially for languages where only a few code samples and tests are (yet) available.
    \item[Coverage guidance.] Learning from greybox fuzzers, we want to integrate coverage guidance into \tool, leveraging LibAFL~\cite{libafl} to guide both input generation and constraint solving towards yet uncovered code.
    \item[Efficient constraint solving.] Overall, \tool is about a million times faster in solving constraints than \isla, which relied on symbolic constraint solving.
    This calls for experiments in which we apply \tool to common constraint-solving benchmarks from the literature, and measure how it fares against other symbolic or heuristic constraint solvers.
    \item[Further optimizations.] We want to explore further optimizations of our generated code.
    This includes leveraging equality saturation~\cite{equality-saturation} to pre-optimize constraints and an analysis of which constraint features favor which optimization algorithms, e.g., particle swarm~\cite{pso}.
    \item[Symbolic constraint solving.]
    We plan to include symbolic constraint solving as a complementary technique to our evolutionary algorithms, notably for efficient solving of constraints involving complex arithmetic expressions or to show that constraints are unsatisfiable.
    \item[Usability.] While \tool is easy to use for Rust programmers, it is less so for non-Rust programmers---and while most developers can be assumed to be familiar with formal grammars, this may be less of a case for experts in the applications being tested.
    We want to explore ways to lower the usability barrier.
    This includes considering domain-specific languages for defining constraints, extracting grammars and constraints from code and input samples, tools for visual editing of grammars and constraints, and training large language models to produce and evolve language specifications from natural language descriptions.
    \item[\tool and \fandango.] We are in contact with the \fandango{} team to exchange ideas and experiences, notably considering algorithms and usability.
    While the \fandango team is planning to incorporate our algorithmic improvements into their Python code, we are considering integrating recent \fandango{} features into \tool, such as support for protocol testing or out-of-grammar mutations.
\end{description}

\begin{acks}
This work was supported in part by the European Research Council (ERC) under the consolidator grant RS$^3$ (101045669).
We would like to thank \href{https://github.com/riesentoaster/}{Valentin Huber} for his implementation of the FANDANGO profiler used throughout our experiments.
In addition, we would like to thank all of the developers and reviewers of the Rust next-generation trait solver and the corresponding rust-analyzer support: \href{https://github.com/lcnr}{Bastian Kauschke}, \href{https://github.com/compiler-errors}{Michael Goulet}, \href{https://github.com/BoxyUwU}{Boxy}, \href{https://github.com/lqd}{R\'emy Rakic}, \href{https://github.com/jackh726}{Jack Huey}, \href{https://github.com/lnicola}{Laurențiu Nicola}, \href{https://github.com/ShoyuVanilla}{Shoyu Vanilla}, \href{https://github.com/ChayimFriedman2}{Chayim Refael Friedman}, and many others, without the work of whom our development process would have been significantly less pleasant.
\end{acks}

\appendix

\section*{Data-Availability Statement} \label{sec:data-availability}

\iftrue
Science is for everyone, and thus we publish our materials.
Our artifact is available online:

\begin{center}
\url{https://github.com/fandango-fuzzer/fandango-rs/}
\end{center}
\else
We commit to publishing our software for further research and review.
Our anonymized artifact is available in HotCRP as supplementary materials and will be published on GitHub pending acceptance, with this section updated accordingly.
\fi

\clearpage

\bibliographystyle{ACM-Reference-Format}
\bibliography{strings,refs}

@article{lampropoulos2019coverage,
  title={Coverage guided, property based testing},
  author={Lampropoulos, Leonidas and Hicks, Michael and Pierce, Benjamin C},
  journal={Proceedings of the ACM on Programming Languages},
  volume={3},
  number={OOPSLA},
  pages={1--29},
  year={2019},
  publisher={ACM New York, NY, USA}
}

@article{claessen2015generating,
  title={Generating constrained random data with uniform distribution},
  author={Claessen, Koen and Dureg{\aa}rd, Jonas and Pa{\l}ka, Micha{\l} H},
  journal={Journal of functional programming},
  volume={25},
  pages={e8},
  year={2015},
  publisher={Cambridge University Press}
}

@inproceedings{reddy2020quickly,
  title={Quickly generating diverse valid test inputs with reinforcement learning},
  author={Reddy, Sameer and Lemieux, Caroline and Padhye, Rohan and Sen, Koushik},
  booktitle={Proceedings of the ACM/IEEE 42nd International Conference on Software Engineering},
  pages={1410--1421},
  year={2020}
}

@article{lampropoulos2017generating,
  title={Generating good generators for inductive relations},
  author={Lampropoulos, Leonidas and Paraskevopoulou, Zoe and Pierce, Benjamin C},
  journal={Proceedings of the ACM on Programming Languages},
  volume={2},
  number={POPL},
  pages={1--30},
  year={2017},
  publisher={ACM New York, NY, USA}
}

@inproceedings{holler2012fuzzing,
  title={Fuzzing with code fragments},
  author={Holler, Christian and Herzig, Kim and Zeller, Andreas},
  booktitle={21st USENIX Security Symposium (USENIX Security 12)},
  pages={445--458},
  year={2012}
}

@inproceedings{srivastava2021gramatron,
  title={Gramatron: Effective grammar-aware fuzzing},
  author={Srivastava, Prashast and Payer, Mathias},
  booktitle={Proceedings of the 30th acm sigsoft international symposium on software testing and analysis},
  pages={244--256},
  year={2021}
}

@article{chen2020survey,
  title={A survey of compiler testing},
  author={Chen, Junjie and Patra, Jibesh and Pradel, Michael and Xiong, Yingfei and Zhang, Hongyu and Hao, Dan and Zhang, Lu},
  journal={ACM Computing Surveys (CSUR)},
  volume={53},
  number={1},
  pages={1--36},
  year={2020},
  publisher={ACM New York, NY, USA}
}

@article{heuristic,
  title={Heuristic and meta-heuristic algorithms and their relevance to the real world: a survey},
  author={Desale, Sachin and Rasool, Akhtar and Andhale, Sushil and Rane, Priti},
  journal={International Journal of Computer Engineering in Research Trends},
  volume={351},
  number={5},
  pages={2349--7084},
  year={2015}
}

@article{Gopinath_Zeller_2019, title={Building Fast Fuzzers}, url={http://arxiv.org/abs/1911.07707}, DOI={10.48550/arXiv.1911.07707}, abstractNote={Fuzzing is one of the key techniques for evaluating the robustness of programs against attacks. Fuzzing has to be effective in producing inputs that cover functionality and ﬁnd vulnerabilities. But it also has to be efﬁcient in producing such inputs quickly. Random fuzzers are very efﬁcient, as they can quickly generate random inputs; but they are not very effective, as the large majority of inputs generated is syntactically invalid. Grammar-based fuzzers make use of a grammar (or another model for the input language) to produce syntactically correct inputs, and thus can quickly cover input space and associated functionality. Existing grammar-based fuzzers are surprisingly inefﬁcient, though: Even the fastest grammar fuzzer dharma still produces inputs about a thousand times slower than the fastest random fuzzer. So far, one can have an effective or an efﬁcient fuzzer, but not both. In this paper, we describe how to build fast grammar fuzzers from the ground up, treating the problem of fuzzing from a programming language implementation perspective. Starting with a Python textbook approach, we adopt and adapt optimization techniques from functional programming and virtual machine implementation techniques together with other novel domain-speciﬁc optimizations in a step-by-step fashion. In our F1 prototype fuzzer, these improve production speed by a factor of 100–300 over the fastest grammar fuzzer dharma. As F1 is even 5–8 times faster than a lexical random fuzzer, we can ﬁnd bugs faster and test with much larger valid inputs than previously possible.}, note={arXiv:1911.07707 [cs]}, number={arXiv:1911.07707}, publisher={arXiv}, author={Gopinath, Rahul and Zeller, Andreas}, year={2019}, month=nov, language={en}, journal={arXiv} }

@article{yarpgen,
author = {Livinskii, Vsevolod and Babokin, Dmitry and Regehr, John},
title = {Random testing for C and C++ compilers with YARPGen},
year = {2020},
issue_date = {November 2020},
publisher = {Association for Computing Machinery},
address = {New York, NY, USA},
volume = {4},
number = {OOPSLA},
url = {https://doi.org/10.1145/3428264},
doi = {10.1145/3428264},
journal = {Proc. ACM Program. Lang.},
month = nov,
articleno = {196},
numpages = {25}
}

@misc{stackover,
  author = {{Stack Exchange Inc.}},
  title = {2024 {Stack Overflow} Developer Survey},
  year = 2024,
  howpublished = {\url{https://survey.stackoverflow.co/2024/}},
}

@misc{csmith-ccc-problems,
  author = {{fuhsnn}},
  title = {Result from fuzzing with csmith and yarpgen},
  year = 2026,
  howpublished={\url{https://github.com/anthropics/claudes-c-compiler/issues/227}}
}

@misc{ccc-compiler,
  author = {{Anthropic}},
  title = {{CCC} — {Claude}'s {C} Compiler},
  year = 2026,
  howpublished = {\url{https://github.com/anthropics/claudes-c-compiler}},
}

@inproceedings{visitorpattern,
  author       = {Jens Palsberg and
                  C. Barry Jay},
  title        = {The Essence of the Visitor Pattern},
  booktitle    = {22nd International Computer Software and Applications Conference},
  pages        = {9--15},
  publisher    = {{IEEE} Computer Society},
  year         = {1998},
  url          = {https://doi.org/10.1109/CMPSAC.1998.716629},
  doi          = {10.1109/CMPSAC.1998.716629},
  bibsource    = {dblp computer science bibliography, https://dblp.org}
}

@incollection{kudelic2022feedback,
  author       = {Robert Kudelic},
  title        = {Feedback Arc Set - {A} History of the Problem and Algorithms},
  series       = {Springer Briefs in Computer Science},
  publisher    = {Springer},
  year         = {2022},
  url          = {https://doi.org/10.1007/978-3-031-10515-9},
  doi          = {10.1007/978-3-031-10515-9},
  isbn         = {978-3-031-10514-2},
  bibsource    = {dblp computer science bibliography, https://dblp.org}
}

@inproceedings{sun2023property,
  title={Property-based fuzzing for finding data manipulation errors in android apps},
  author={Sun, Jingling and Su, Ting and Jiang, Jiayi and Wang, Jue and Pu, Geguang and Su, Zhendong},
  booktitle={Proceedings of the 31st ACM joint european software engineering conference and symposium on the foundations of software engineering},
  pages={1088--1100},
  year={2023}
}

@inproceedings{xie2022rozz,
  title={ROZZ: property-based fuzzing for robotic programs in ROS},
  author={Xie, Kai-Tao and Bai, Jia-Ju and Zou, Yong-Hao and Wang, Yu-Ping},
  booktitle={2022 International Conference on Robotics and Automation (ICRA)},
  pages={6786--6792},
  year={2022},
  organization={IEEE}
}

@inproceedings{zang2022compiler,
  title={Compiler testing using template java programs},
  author={Zang, Zhiqiang and Wiatrek, Nathan and Gligoric, Milos and Shi, August},
  booktitle={Proceedings of the 37th IEEE/ACM International Conference on Automated Software Engineering},
  pages={1--13},
  year={2022}
}

@inproceedings{sharma2023rustsmith,
  title={Rustsmith: Random differential compiler testing for rust},
  author={Sharma, Mayank and Yu, Pingshi and Donaldson, Alastair F},
  booktitle={Proceedings of the 32nd ACM SIGSOFT International Symposium on Software Testing and Analysis},
  pages={1483--1486},
  year={2023}
}

@article{korel1992dynamic,
  title={Dynamic method for software test data generation},
  author={Korel, Bogdan},
  journal={Software Testing, Verification and Reliability},
  volume={2},
  number={4},
  pages={203--213},
  year={1992},
  publisher={Wiley Online Library}
}

@article{segura2016survey,
  title={A survey on metamorphic testing},
  author={Segura, Sergio and Fraser, Gordon and Sanchez, Ana B and Ruiz-Cort{\'e}s, Antonio},
  journal={IEEE Transactions on software engineering},
  volume={42},
  number={9},
  pages={805--824},
  year={2016},
  publisher={IEEE}
}

@article{chen2018metamorphic,
  title={Metamorphic testing: A review of challenges and opportunities},
  author={Chen, Tsong Yueh and Kuo, Fei-Ching and Liu, Huai and Poon, Pak-Lok and Towey, Dave and Tse, TH and Zhou, Zhi Quan},
  journal={ACM Computing Surveys (CSUR)},
  volume={51},
  number={1},
  pages={1--27},
  year={2018},
  publisher={ACM New York, NY, USA}
}

@inproceedings{malburg2011combining,
  title={Combining search-based and constraint-based testing},
  author={Malburg, Jan and Fraser, Gordon},
  booktitle={2011 26th IEEE/ACM International Conference on Automated Software Engineering (ASE 2011)},
  pages={436--439},
  year={2011},
  organization={IEEE}
}

@inproceedings{zhu2019improving,
  title={Improving search-based software testing by constraint-based genetic operators},
  author={Zhu, Ziming and Jiao, Li},
  booktitle={Proceedings of the Genetic and Evolutionary Computation Conference},
  pages={1435--1442},
  year={2019}
}

@inproceedings{nguyen2022bedivfuzz,
  title={Bedivfuzz: Integrating behavioral diversity into generator-based fuzzing},
  author={Nguyen, Hoang Lam and Grunske, Lars},
  booktitle={Proceedings of the 44th international conference on software engineering},
  pages={249--261},
  year={2022}
}

@article{boyapati2002korat,
  title={Korat: Automated testing based on Java predicates},
  author={Boyapati, Chandrasekhar and Khurshid, Sarfraz and Marinov, Darko},
  journal={ACM SIGSOFT Software Engineering Notes},
  volume={27},
  number={4},
  pages={123--133},
  year={2002},
  publisher={ACM New York, NY, USA}
}

@inproceedings{albunian2020measuring,
  title={Measuring and maintaining population diversity in search-based unit test generation},
  author={Albunian, Nasser and Fraser, Gordon and Sudholt, Dirk},
  booktitle={International Symposium on Search Based Software Engineering},
  pages={153--168},
  year={2020},
  organization={Springer}
}

@inproceedings{albunian2017diversity,
  title={Diversity in search-based unit test suite generation},
  author={Albunian, Nasser M},
  booktitle={International Symposium on Search Based Software Engineering},
  pages={183--189},
  year={2017},
  organization={Springer}
}

@inproceedings{xanthakis1992application,
  title={Application of genetic algorithms to software testing},
  author={Xanthakis, Spiros and Ellis, C and Skourlas, Christos and Le Gall, A and Katsikas, S and Karapoulios, K},
  booktitle={Proceedings of the 5th International Conference on Software Engineering and Applications},
  pages={625--636},
  year={1992}
}

@article{korel1990automated,
  title={Automated software test data generation},
  author={Korel, Bogdan},
  journal={IEEE Transactions on software engineering},
  volume={16},
  number={8},
  pages={870--879},
  year={1990},
  publisher={IEEE}
}

@article{miller1976automatic,
  title={Automatic generation of floating-point test data},
  author={Miller, Webb and Spooner, David L.},
  journal={IEEE Transactions on Software Engineering},
  number={3},
  pages={223--226},
  year={1976},
  publisher={IEEE}
}

@inproceedings{goldstein2024property,
  title={Property-based testing in practice},
  author={Goldstein, Harrison and Cutler, Joseph W and Dickstein, Daniel and Pierce, Benjamin C and Head, Andrew},
  booktitle={Proceedings of the IEEE/ACM 46th International Conference on Software Engineering},
  pages={1--13},
  year={2024}
}

@inproceedings{yatoh2014arbitcheck,
  title={Arbitcheck: A highly automated property-based testing tool for java},
  author={Yatoh, Kohsuke and Sakamoto, Kazunori and Ishikawa, Fuyuki and Honiden, Shinichi},
  booktitle={2014 IEEE Seventh International Conference on Software Testing, Verification and Validation Workshops},
  pages={405--412},
  year={2014},
  organization={IEEE}
}

@inproceedings{padhye2019jqf,
  title={Jqf: Coverage-guided property-based testing in java},
  author={Padhye, Rohan and Lemieux, Caroline and Sen, Koushik},
  booktitle={Proceedings of the 28th ACM SIGSOFT International Symposium on Software Testing and Analysis},
  pages={398--401},
  year={2019}
}

@inproceedings{gu2023llm,
  title={Llm-based code generation method for golang compiler testing},
  author={Gu, Qiuhan},
  booktitle={Proceedings of the 31st ACM Joint European Software Engineering Conference and Symposium on the Foundations of Software Engineering},
  pages={2201--2203},
  year={2023}
}

@article{yang2024whitefox,
  title={Whitefox: White-box compiler fuzzing empowered by large language models},
  author={Yang, Chenyuan and Deng, Yinlin and Lu, Runyu and Yao, Jiayi and Liu, Jiawei and Jabbarvand, Reyhaneh and Zhang, Lingming},
  journal={Proceedings of the ACM on Programming Languages},
  volume={8},
  number={OOPSLA2},
  pages={709--735},
  year={2024},
  publisher={ACM New York, NY, USA}
}

@inproceedings{ye2023generative,
  title={A generative and mutational approach for synthesizing bug-exposing test cases to guide compiler fuzzing},
  author={Ye, Guixin and Hu, Tianmin and Tang, Zhanyong and Fan, Zhenye and Tan, Shin Hwei and Zhang, Bo and Qian, Wenxiang and Wang, Zheng},
  booktitle={Proceedings of the 31st ACM Joint European Software Engineering Conference and Symposium on the Foundations of Software Engineering},
  pages={1127--1139},
  year={2023}
}

@article{even2022csmithedge,
  title={CsmithEdge: more effective compiler testing by handling undefined behaviour less conservatively},
  author={Even-Mendoza, Karine and Cadar, Cristian and Donaldson, Alastair F},
  journal={Empirical Software Engineering},
  volume={27},
  number={6},
  pages={129},
  year={2022},
  publisher={Springer}
}

@inproceedings{chaliasos2022finding,
  title={Finding typing compiler bugs},
  author={Chaliasos, Stefanos and Sotiropoulos, Thodoris and Spinellis, Diomidis and Gervais, Arthur and Livshits, Benjamin and Mitropoulos, Dimitris},
  booktitle={Proceedings of the 43rd ACM SIGPLAN International Conference on Programming Language Design and Implementation},
  pages={183--198},
  year={2022}
}

@incollection{breuer1971algorithm,
  title={An algorithm for generating a fault detection test for a class of sequential circuits},
  author={Breuer, Melvin A},
  booktitle={Theory of Machines and Computations},
  pages={313--326},
  year={1971},
  publisher={Elsevier}
}

@article{miller1990empirical,
  title={An empirical study of the reliability of UNIX utilities},
  author={Miller, Barton P and Fredriksen, Lars and So, Bryan},
  journal={Communications of the ACM},
  volume={33},
  number={12},
  pages={32--44},
  year={1990},
  publisher={ACM New York, NY, USA}
}

@misc{RAII,
  key = {RAII},
  title = {Resource Acquisition Is Initialization},
  howpublished = {\url{https://doc.rust-lang.org/rust-by-example/scope/raii.html}},
  note = {Retrieved 2025-31-05}
}

@misc{folklore,
  title = {folklore in nLab},
  howpublished = {\url{https://ncatlab.org/nlab/show/folklore}},
  note = {Retrieved 2026-13-03},
  year = 2023,
  author = {Urs Schreiber}
}

@techreport{dispatcher,
  title={Dynamic Dispatch in Object-Oriented Languages},
  author={Milton, Scott and Schmidt, Heinz W.},
  year={1994},
  institution={Australian National University}
}

@misc{Falk_2025, type={Rust}, title={gamozolabs/fzero\_fuzzer}, rights={MIT}, url={https://github.com/gamozolabs/fzero\_fuzzer}, abstractNote={A fast Rust-based safe and thread-friendly grammar-based fuzz generator}, author={Falk, Brandon}, year={2019}, month=nov, publisher = {GitHub}, journal = {GitHub repository} }

@misc{blazingly-fast, title={r/rustjerk - {Anything built in Rust is blazingly fast by default}}, url={https://www.reddit.com/r/rustjerk/comments/sqy221/anything_built_in_rust_is_blazingly_fast_by/}, author={solidiquis1}, year={2019}, month=nov, publisher = {Reddit}, journal = {Reddit post} }

@article{Chomsky_1956, title={Three models for the description of language}, volume={2}, rights={https://ieeexplore.ieee.org/Xplorehelp/downloads/license-information/IEEE.html}, ISSN={0018-9448}, DOI={10.1109/TIT.1956.1056813}, number={3}, journal={IEEE Transactions on Information Theory}, author={Chomsky, N.}, year={1956}, month=sep, pages={113–124}, language={en} }

@inproceedings{aschermann2019nautilus,
  author       = {Cornelius Aschermann and
                  Tommaso Frassetto and
                  Thorsten Holz and
                  Patrick Jauernig and
                  Ahmad{-}Reza Sadeghi and
                  Daniel Teuchert},
  title        = {{NAUTILUS:} Fishing for Deep Bugs with Grammars},
  booktitle    = {26th Annual Network and Distributed System Security Symposium},
  publisher    = {The Internet Society},
  year         = {2019},
  url          = {https://www.ndss-symposium.org/ndss-paper/nautilus-fishing-for-deep-bugs-with-grammars/},
  bibsource    = {dblp computer science bibliography, https://dblp.org}
}

@article{rentsch1982object,
  author       = {Tim Rentsch},
  title        = {Object Oriented Programming},
  journal      = {{ACM} {SIGPLAN} Notices},
  volume       = {17},
  number       = {9},
  pages        = {51--57},
  year         = {1982},
  url          = {https://doi.org/10.1145/947955.947961},
  doi          = {10.1145/947955.947961},
  bibsource    = {dblp computer science bibliography, https://dblp.org}
}

@article{russell2001events,
  author       = {George Russell},
  title        = {Events in Haskell, and How to Implement Them},
  booktitle    = {Proceedings of the Sixth {ACM} {SIGPLAN} International Conference
                  on Functional Programming},
  pages        = {157--168},
  publisher    = {{ACM}},
  year         = {2001},
  url          = {https://doi.org/10.1145/507635.507655},
  doi          = {10.1145/507635.507655},
  bibsource    = {dblp computer science bibliography, https://dblp.org}
}

@inproceedings{hodovan2018grammarinator,
  author       = {Ren{\'{a}}ta Hodov{\'{a}}n and
                  {\'{A}}kos Kiss and
                  Tibor Gyim{\'{o}}thy},
  title        = {Grammarinator: a grammar-based open source fuzzer},
  booktitle    = {Proceedings of the 9th {ACM} {SIGSOFT} International Workshop on Automating
                  {TEST} Case Design},
  pages        = {45--48},
  publisher    = {{ACM}},
  year         = {2018},
  url          = {https://doi.org/10.1145/3278186.3278193},
  doi          = {10.1145/3278186.3278193},
  bibsource    = {dblp computer science bibliography, https://dblp.org}
}

@inproceedings{eberlein2020evolutionary,
  title={Evolutionary grammar-based fuzzing},
  author={Eberlein, Martin and Noller, Yannic and Vogel, Thomas and Grunske, Lars},
  booktitle={Search-Based Software Engineering: 12th International Symposium, SSBSE 2020, Bari, Italy, October 7--8, 2020, Proceedings 12},
  pages={105--120},
  year={2020},
  organization={Springer}
}

@inproceedings{godefroid2008grammar,
  author       = {Patrice Godefroid and
                  Adam Kiezun and
                  Michael Y. Levin},
  title        = {Grammar-based whitebox fuzzing},
  booktitle    = {Proceedings of the {ACM} {SIGPLAN} 2008 Conference on Programming
                  Language Design and Implementation, Tucson, AZ, USA, June 7-13, 2008},
  pages        = {206--215},
  publisher    = {{ACM}},
  year         = {2008},
  url          = {https://doi.org/10.1145/1375581.1375607},
  doi          = {10.1145/1375581.1375607},
  bibsource    = {dblp computer science bibliography, https://dblp.org}
}

@inproceedings{havrikov2019systematically,
  author       = {Nikolas Havrikov and
                  Andreas Zeller},
  title        = {Systematically Covering Input Structure},
  booktitle    = {34th {IEEE/ACM} International Conference on Automated Software Engineering},
  pages        = {189--199},
  publisher    = {{IEEE}},
  year         = {2019},
  url          = {https://doi.org/10.1109/ASE.2019.00027},
  doi          = {10.1109/ASE.2019.00027},
  bibsource    = {dblp computer science bibliography, https://dblp.org}
}

@inproceedings{zamudio2025fandango,
  author = {Zamudio Amaya, Jos\'{e} Antonio and Smytzek, Marius and Zeller, Andreas},
  booktitle = {Proceedings of the 34th ACM SIGSOFT International Symposium on Software Testing and Analysis},
  title = {{FANDANGO}: {E}volving Language-Based Testing},
  year = {2025},
  address = {Trondheim, Norway},
  publisher = {ACM},
  url = {https://doi.org/10.1145/3728915}
}

@misc{libfuzzer,
  author = {{LLVM Project}},
  title = {{libFuzzer}},
  howpublished = {\url{https://llvm.org/docs/LibFuzzer.html}},
  note = {Retrieved 2025-10-08}
}

@inproceedings{yang2011csmith,
 author = {Yang, Xuejun and Chen, Yang and Eide, Eric and Regehr, John},
 title = {Finding and Understanding Bugs in {C} Compilers},
 booktitle = pldi,
 series = {PLDI '11},
 year = {2011},
 isbn = {978-1-4503-0663-8},
 location = {San Jose, California, USA},
 pages = {283--294},
 numpages = {12},
 url = {http://doi.acm.org/10.1145/1993498.1993532},
 doi = {10.1145/1993498.1993532},
 acmid = {1993532},
 publisher = {ACM},
 PUBLISHERaddress = {New York, NY, USA},
}

@software{ghidra,
author = {{National Security Agency}},
license = {Apache-2.0},
month = aug,
title = {{Ghidra}},
url = {https://github.com/NationalSecurityAgency/ghidra},
version = {11.4.2},
year = {2025}
}

@misc{aflfuzz,
  ids = {AFLFuzz},
  author = {Michal Za\l{}ewski},
  title = {American Fuzzy Lop},
  year = {2018},
  howpublished = {\url{http://lcamtuf.coredump.cx/afl/}},
  note = {Accessed: 2025-10-10}
}

@book{rustbook, address={New York}, title={The Rust Programming Language, 2nd Edition}, ISBN={978-1-7185-0310-6}, publisher={No Starch Press}, author={Klabnik, Steve and Nichols, Carol}, year={2023}, language={eng} }

@article{dominicACM,
  author       = {Dominic Steinh{\"{o}}fel and
                  Andreas Zeller},
  title        = {Language-Based Software Testing},
  journal      = {Communications of the {ACM}},
  volume       = {67},
  number       = {4},
  pages        = {80--84},
  year         = {2024},
  url          = {https://doi.org/10.1145/3631520},
  doi          = {10.1145/3631520},
  bibsource    = {dblp computer science bibliography, https://dblp.org}
}

@inproceedings{steinhofel2022input,
  author       = {Dominic Steinh{\"{o}}fel and
                  Andreas Zeller},
  title        = {Input Invariants},
  booktitle    = {Software Engineering 2023, Fachtagung des GI-Fachbereichs Softwaretechnik},
  series       = {{LNI}},
  volume       = {{P-332}},
  pages        = {113--114},
  publisher    = {Gesellschaft f{\"{u}}r Informatik e.V.},
  year         = {2023},
  url          = {https://dl.gi.de/handle/20.500.12116/40110},
  bibsource    = {dblp computer science bibliography, https://dblp.org}
}

@inproceedings{wang2019superion,
  author       = {Junjie Wang and
                  Bihuan Chen and
                  Lei Wei and
                  Yang Liu},
  title        = {Superion: grammar-aware greybox fuzzing},
  booktitle    = {Proceedings of the 41st International Conference on Software Engineering},
  pages        = {724--735},
  publisher    = {{IEEE} / {ACM}},
  year         = {2019},
  url          = {https://doi.org/10.1109/ICSE.2019.00081},
  doi          = {10.1109/ICSE.2019.00081},
  bibsource    = {dblp computer science bibliography, https://dblp.org}
}

@inproceedings{bendrissou2023grammar,
  author       = {Bachir Bendrissou and
                  Cristian Cadar and
                  Alastair F. Donaldson},
  title        = {Grammar Mutation for Testing Input Parsers (Registered Report)},
  booktitle    = {Proceedings of the 2nd International Fuzzing Workshop},
  pages        = {3--11},
  publisher    = {{ACM}},
  year         = {2023},
  url          = {https://doi.org/10.1145/3605157.3605170},
  doi          = {10.1145/3605157.3605170},
  bibsource    = {dblp computer science bibliography, https://dblp.org}
}

@Article{burkhardt1967,
  author       = {Walter H. Burkhardt},
  title        = {Generating test programs from syntax},
  journal      = {Computing},
  volume       = {2},
  number       = {1},
  pages        = {53--73},
  year         = {1967},
  url          = {https://doi.org/10.1007/BF02235512},
  doi          = {10.1007/BF02235512},
  bibsource    = {dblp computer science bibliography, https://dblp.org}
}

@inproceedings{claessen2000quickcheck,
  title={QuickCheck: a lightweight tool for random testing of Haskell programs},
  author={Claessen, Koen and Hughes, John},
  booktitle={Proceedings of the fifth ACM SIGPLAN international conference on Functional programming},
  pages={268--279},
  year={2000}
}

@article{hanford1970,
  title={Automatic generation of test cases},
  author={Hanford, Kenneth V.},
  journal={IBM Systems Journal},
  volume={9},
  number={4},
  pages={242--257},
  year={1970},
  publisher={IBM}
}

@article{purdom1972,
  author={Purdom, Paul},
  title={A sentence generator for testing parsers},
  year={1972},
  issn={0006-3835},
  journal={BIT Numerical Mathematics},
  volume={12},
  number={3},
  doi={10.1007/BF01932308},
  url={http://dx.doi.org/10.1007/BF01932308},
  publisher={Kluwer Academic Publishers},
  pages={366-375},
  language={English}
}

@InProceedings{demoura2008z3,
  author       = {Leonardo Mendon{\c{c}}a de Moura and
                  Nikolaj S. Bj{\o}rner},
  editor       = {C. R. Ramakrishnan and
                  Jakob Rehof},
  title        = {{Z3:} An Efficient {SMT} Solver},
  booktitle    = {14th International Conference on Tools and Algorithms for the Construction and Analysis of Systems},
  series       = {Lecture Notes in Computer Science},
  volume       = {4963},
  pages        = {337--340},
  publisher    = {Springer},
  year         = {2008},
  url          = {https://doi.org/10.1007/978-3-540-78800-3\_24},
  doi          = {10.1007/978-3-540-78800-3\_24},
  bibsource    = {dblp computer science bibliography, https://dblp.org}
}

@article{enumerating-cfgs, title={How to enumerate trees from a context-free grammar}, url={http://arxiv.org/abs/2305.00522}, DOI={10.48550/arXiv.2305.00522}, journal={arXiv}, author={Piantadosi, Steven T.}, year={2023}, month=apr, language={en} }

@article{nsga2,
  author       = {Kalyanmoy Deb and
                  Samir Agrawal and
                  Amrit Pratap and
                  T. Meyarivan},
  title        = {A fast and elitist multiobjective genetic algorithm: {NSGA-II}},
  journal      = {{IEEE} Transactions on Evolutionary Computing},
  volume       = {6},
  number       = {2},
  pages        = {182--197},
  year         = {2002},
  url          = {https://doi.org/10.1109/4235.996017},
  doi          = {10.1109/4235.996017},
  bibsource    = {dblp computer science bibliography, https://dblp.org}
}

@book{mahfoud1995niching,
  title={Niching methods for genetic algorithms},
  author={Mahfoud, Samir W},
  year={1995},
  publisher={University of Illinois at Urbana-Champaign}
}

@article{weighted-sum-moea, title={The weighted sum method for multi-objective optimization: new insights}, volume={41}, ISSN={1615-1488}, DOI={10.1007/s00158-009-0460-7}, abstractNote={As a common concept in multi-objective optimization, minimizing a weighted sum constitutes an independent method as well as a component of other methods. Consequently, insight into characteristics of the weighted sum method has far reaching implications. However, despite the many published applications for this method and the literature addressing its pitfalls with respect to depicting the Pareto optimal set, there is little comprehensive discussion concerning the conceptual significance of the weights and techniques for maximizing the effectiveness of the method with respect to a priori articulation of preferences. Thus, in this paper, we investigate the fundamental significance of the weights in terms of preferences, the Pareto optimal set, and objective-function values. We determine the factors that dictate which solution point results from a particular set of weights. Fundamental deficiencies are identified in terms of a priori articulation of preferences, and guidelines are provided to help avoid blind use of the method.}, number={6}, journal={Structural and Multidisciplinary Optimization}, author={Marler, R. Timothy and Arora, Jasbir S.}, year={2010}, month=jun, pages={853–862}, language={en} }

@article{lexicographic-optimization, title={Lexicographic quasiconcave multiobjective programming}, volume={21}, ISSN={1432-5217}, DOI={10.1007/BF01919766}, abstractNote={Letfi:A → R ben real-valued objective functions on a convex setA ⊂-Km,K:=R orC, n, m∈N. Letg: A → Rnbe defined by$$g(x): = (g_1 (x),...,g_n (x)): = (f_{i_1 (x)} (x),...,f_{i_n (x)} (x))$$, where for eachx∈A, (i1(x), ..., in(x)) is a permutation of (1, ...,n) such that$$(f_{i_1 (x)} (x) leqslant ... leqslant f_{i_n (x)} (x))$$. In this paper we treat the problem of findingx*∈A such that$$g(x^ *  ) = mathop {l--max }limits_{x in A} g(x)$$, wherel-max denotes the lexicographic maximum. If the fi’s are strongly quasiconcave we can reduce the problem stepwise until finally it is in the form of a scalar programming problem. Further, we consider conditions for the existence and uniqueness of a solution and discuss the relationship of the problem to the vector maximum (i.e. Pareto) and maxmin (i.e. Chebychev) problems.}, number={3}, journal={Zeitschrift für Operations Research}, author={Behringer, F. A.}, year={1977}, month=jun, pages={103–116}, language={en} }

@article{pareto-front-definition, title={Genetic algorithms: Foundations and applications}, volume={21}, ISSN={1572-9338}, DOI={10.1007/BF02022092}, abstractNote={Genetic algorithms are defined. Attention is directed to why they work: schemas and building blocks, implicit parallelism, and exponentially biased sampling of the better schema. Why they fail and how undesirable behavior can be overcome is discussed. Current genetic algorithm practice is summarized. Five successful applications are illustrated: image registration, AEGIS surveillance, network configuration, prisoner’s dilemma, and gas pipeline control. Three classes of problems for which genetic algorithms are ill suited are illustrated: ordering problems, smooth optimization problems, and “totally indecomposable” problems.}, number={1}, journal={Annals of Operations Research}, author={Liepins, G. E. and Hilliard, M. R.}, year={1989}, month=dec, pages={31–57}, language={en} }

@inproceedings{novelty-search,
  author       = {Joel Lehman and
                  Kenneth O. Stanley},
  title        = {Exploiting Open-Endedness to Solve Problems Through the Search for
                  Novelty},
  booktitle    = {Proceedings of the Eleventh International Conference on the Synthesis and Simulation of Living Systems},
  pages        = {329--336},
  publisher    = {{MIT} Press},
  year         = {2008},
  url          = {http://mitpress2.mit.edu/books/chapters/0262287196chap43.pdf},
  bibsource    = {dblp computer science bibliography, https://dblp.org}
}

@article{novelty-fuzzing,
  author       = {Nico Schiller and
                  Xinyi Xu and
                  Lukas Bernhard and
                  Nils Bars and
                  Moritz Schloegel and
                  Thorsten Holz},
  title        = {Novelty Not Found: Exploring Input Shadowing in Fuzzing through Adaptive
                  Fuzzer Restarts},
  journal      = {{ACM} Transactions on Software Engineering Methodologies},
  volume       = {34},
  number       = {3},
  pages        = {85:1--85:32},
  year         = {2025},
  url          = {https://doi.org/10.1145/3712186},
  doi          = {10.1145/3712186},
  bibsource    = {dblp computer science bibliography, https://dblp.org}
}

@inproceedings{pso,
  author       = {Konstantinos E. Parsopoulos and
                  Michael N. Vrahatis},
  title        = {Particle swarm optimization method in multiobjective problems},
  booktitle    = {Proceedings of the 2002 {ACM} Symposium on Applied Computing},
  pages        = {603--607},
  publisher    = {{ACM}},
  year         = {2002},
  url          = {https://doi.org/10.1145/508791.508907},
  doi          = {10.1145/508791.508907},
  bibsource    = {dblp computer science bibliography, https://dblp.org}
}

@article{equality-saturation,
  author       = {Max Willsey and
                  Chandrakana Nandi and
                  Yisu Remy Wang and
                  Oliver Flatt and
                  Zachary Tatlock and
                  Pavel Panchekha},
  title        = {egg: Fast and extensible equality saturation},
  journal      = {Proceedings of the ACM on Programming Languages},
  volume       = {5},
  number       = {{POPL}},
  pages        = {1--29},
  year         = {2021},
  url          = {https://doi.org/10.1145/3434304},
  doi          = {10.1145/3434304},
  bibsource    = {dblp computer science bibliography, https://dblp.org}
}

@misc{monomorphization, author = {{Rustc Dev Guide Working Area}}, title={Monomorphization}, booktitle = {Rust Compiler Development Guide}, url={https://rustc-dev-guide.rust-lang.org/backend/monomorph.html}, year=2025 }

@inproceedings{libafl,
 author       = {Andrea Fioraldi and Dominik Maier and Dongjia Zhang and Davide Balzarotti},
 title        = {{LibAFL: A Framework to Build Modular and Reusable Fuzzers}},
 booktitle    = {Proceedings of the 29th ACM conference on Computer and communications security (CCS)},
 series       = {CCS '22},
 year         = {2022},
 month        = {November},
 location     = {Los Angeles, U.S.A.},
 publisher    = {ACM},
}

@article{coupon-collector,
  author       = {Philippe Flajolet and
                  Dani{\`{e}}le Gardy and
                  Lo{\"{y}}s Thimonier},
  title        = {Birthday Paradox, Coupon Collectors, Caching Algorithms and Self-Organizing
                  Search},
  journal      = {Discrete Applied Mathematics},
  volume       = {39},
  number       = {3},
  pages        = {207--229},
  year         = {1992},
  url          = {https://doi.org/10.1016/0166-218X(92)90177-C},
  doi          = {10.1016/0166-218X(92)90177-C},
  bibsource    = {dblp computer science bibliography, https://dblp.org}
}

@misc{gcc,
  title        = {{GCC, the GNU Compiler Collection}},
  howpublished = {\url{https://gcc.gnu.org/}},
  author       = {{Free Software Foundation}},
  year         = {2025}
}

@misc{next-solver, title={Tracking Issue for `-Znext-solver`}, url={https://github.com/rust-lang/rust/issues/107374}, year={2023}, month=jan, publisher = {GitHub}, journal = {GitHub issue}, author={Kauschke, Bastian} }

@inproceedings{pathafl,
  author       = {Shengbo Yan and
                  Chenlu Wu and
                  Hang Li and
                  Wei Shao and
                  Chunfu Jia},
  title        = {PathAFL: Path-Coverage Assisted Fuzzing},
  booktitle    = {{ASIA} {CCS} '20: The 15th {ACM} Asia Conference on Computer and Communications
                  Security, Taipei, Taiwan, October 5-9, 2020},
  pages        = {598--609},
  publisher    = {{ACM}},
  year         = {2020},
  url          = {https://doi.org/10.1145/3320269.3384736},
  doi          = {10.1145/3320269.3384736},
  bibsource    = {dblp computer science bibliography, https://dblp.org}
}

@inproceedings{prudent-practices,
  author       = {Moritz Schloegel and
                  Nils Bars and
                  Nico Schiller and
                  Lukas Bernhard and
                  Tobias Scharnowski and
                  Addison Crump and
                  Arash Ale Ebrahim and
                  Nicolai Bissantz and
                  Marius Muench and
                  Thorsten Holz},
  title        = {SoK: Prudent Evaluation Practices for Fuzzing},
  booktitle    = {{IEEE} Symposium on Security and Privacy, {SP} 2024, San Francisco,
                  CA, USA, May 19-23, 2024},
  pages        = {1974--1993},
  publisher    = {{IEEE}},
  year         = {2024},
  url          = {https://doi.org/10.1109/SP54263.2024.00137},
  doi          = {10.1109/SP54263.2024.00137},
  bibsource    = {dblp computer science bibliography, https://dblp.org}
}

@inproceedings{path-fuzzer,
  author       = {Giacomo Priamo and
                  Daniele Cono D'Elia and
                  Mathias Payer and
                  Leonardo Querzoni},
  title        = {Towards Path-Aware Coverage-Guided Fuzzing},
  booktitle    = {{IEEE/ACM} International Symposium on Code Generation and Optimization,
                  {CGO} 2026, Sydney, Australia, January 31 - Feb. 4, 2026},
  pages        = {84--97},
  publisher    = {{IEEE}},
  year         = {2026},
  url          = {https://doi.org/10.1109/CGO68049.2026.11395191},
  doi          = {10.1109/CGO68049.2026.11395191},
  bibsource    = {dblp computer science bibliography, https://dblp.org}
}

@misc{crump39c3, address={Hamburg, Germany}, type={Lecture}, title={Demystifying Fuzzer Behaviour}, url={https://media.ccc.de/v/39c3-demystifying-fuzzer-behaviour}, author={Crump, Addison}, year={2025}, month=dec, language={English}, publisher={Chaos Computer Club}, booktitle={39th Chaos Computer Conference} }

@inproceedings{ijon,
  author       = {Cornelius Aschermann and
                  Sergej Schumilo and
                  Ali Abbasi and
                  Thorsten Holz},
  title        = {Ijon: Exploring Deep State Spaces via Fuzzing},
  booktitle    = {2020 {IEEE} Symposium on Security and Privacy, {SP} 2020, San Francisco,
                  CA, USA, May 18-21, 2020},
  pages        = {1597--1612},
  publisher    = {{IEEE}},
  year         = {2020},
  url          = {https://doi.org/10.1109/SP40000.2020.00117},
  doi          = {10.1109/SP40000.2020.00117},
  bibsource    = {dblp computer science bibliography, https://dblp.org}
}

@String{BIT = "{BIT}" }

@String{Computing = "Computing" }

@String{Computer = "{IEEE} Computer" }

@String{Academic = "Academic Press" }

@String{Springer = "Springer-Verlag" }

@String{IEEE = "Institute of Electrical and Electronics Engineers" }

@String{Usenix = "The advanced computing systems association" }

@string{oopsla          = "ACM SIGPLAN Conference on Object-Oriented Programming Systems, Languages, and Applications (OOPSLA)"}

@string{pldi            = "ACM SIGPLAN Conference on Programming Language Design and Implementation (PLDI)"}

@string{ase             = "ACM/IEEE International Conference on Automated Software Engineering (ASE)"}

@string{cgo = "IEEE/ACM International Symposium on Code Generation and Optimization (CGO)"}

@string{fuzzing     = "International Fuzzing Workshop (FUZZING)"}

\appendix

\end{document}